\documentclass[pre,a4paper]{revtex4-1}
\usepackage{amssymb}
\usepackage{amsmath}
\usepackage[T1]{fontenc}
\usepackage[utf8]{inputenc}

\newcommand{\pa}{\partial}

\newcommand{\myref}[1]{(\ref{#1})}

\newcommand{\Om}{\Omega}
\newcommand{\de}{\delta}
\newcommand{\De}{\Delta}
\newcommand{\al}{\alpha}

\newcommand{\la}{\lambda}

\newcommand{\ga}{\gamma}

\newcommand{\be}{\beta}
\newcommand{\te}{\theta}

\newcommand{\sig}{\sigma}

\renewcommand{\leq}{\leqslant}
\renewcommand{\geq}{\geqslant}
\renewcommand{\tilde}[1]{\widetilde{#1}}
\renewcommand{\hat}[1]{\widehat{#1}}
\newcommand{\lan}{\langle}
\newcommand{\ran}{\rangle}
\newcommand{\nab}{\nabla}

\newcommand{\demi}{\frac{1}{2}}

\newcommand{\dpar}[3]{\left(\frac{\partial #1}{\partial #2}\right)_{#3}}

\newlength{\somme}
\newlength{\sommep}
\newcommand{\intvide}{\rule[-\sommep]{0cm}{\somme}}

\newlength{\sommebis}
\newlength{\sommepbis}
\usepackage[T1]{fontenc}
\usepackage{lmodern}
\usepackage{bm,color}
\usepackage{amssymb,amsmath,graphicx}
\usepackage[colorinlistoftodos]{todonotes}
\begin{document}

\title{Voronoi glass-forming liquids : A structural study}
\date{\today}
\author{C. Ruscher}
\author{J. Baschnagel}
\author{J. Farago}
\email[contact:]{jean.farago@ics-cnrs.unistra.fr}
\affiliation{Institut Charles Sadron, Universit\'e de Strasbourg, CNRS UPR 22, Strasbourg, France}


\begin{abstract}
  {We introduce a new theoretical model of simple fluid, whose interactions, defined in terms of the Voronoi cells of the configurations, are local and many-body. The resulting system is studied both theoretically and numerically. We show that the fluid, though sharing the global features of other models of fluids with soft interactions, has several unusual characteristics, which are investigated and discussed.}
\end{abstract}

\maketitle

\section{Introduction}

In the realm of liquid physics, the glass transition remains for the theoretist a fascinating challenge. Despite decades of efforts and a diversity of approaches, no unifying picture has yet emerged, which would be capable of describing successfully the link between structure and dynamics, interactions and slowing down, etc\ldots, in  the whole temperature range and all dynamical regimes covered by the glass transition \cite{CavagnaReport,NieuwenhuizenBook,BerthierBiroliRMP}.

In addition to the dramatic slowing down of the dynamics over a modest temperature range, the glass transition is characterized by some general features, dynamic heterogeneity, activation, enhanced sensitivity to the potential energy landscape on cooling, etc\ldots which are found generally in any model of glass-forming liquid. Some generic properties  common to all fluids or models of fluid considered so far generate qualitatively the same phenomenology, with only quantitative differences one is still unable to account for. A possible route of research consists in considering  models of fluid quite different from the existing ones, and see to what extent the usual glass transition scenario is or is not modified by the modifications introduced in the interactions. For instance, all models considered so far have a (hard or soft) core repulsion associated with a well-defined characteristic length, and usually interact {\em via} pair potentials.

To investigate the sensitivity of the glass transition phenomenology to these particulars, we present in this paper a new model of glass-forming fluid, whose interactions are defined geometrically in terms of the Laguerre-Voronoi tessellations of the configurations. This model, which is the binary version of the so-called Voronoi liquid presented elsewhere \cite{voronoiliquid}, is characterized by interactions that are intrinsically many-body, and by the absence of a proper hare core repulsion.

The (monodisperse) Voronoi liquid  \cite{voronoiliquid} has shown convincingly that this class of fluids, though perfectly thermodynamically stable {\em bona fide} fluids, is able to display strongly unusual traits, for  instance a large mesoscopic range where the sound attenuation behaves anomalously. It has been shown in  \cite{papersound} that this regime does not violate the existing body of knowledge on liquid physics \cite{Balucani}, but that the atypical definition  of the Voronoi fluid induces atypical numerical values of the basic structural and thermodynamic observables, allowing the fluid to visit some ``regions'' of the parameter space which are never probed by more widespread models, and thus to reveal this anomalous regime.

As a result, the atypicality of the fluid has been successfully used to probe the liquid state and get from it a better understanding of the relation between structure and dynamics. To probe similarly the supercooled regime, the Voronoi liquid is not adapted, since it crystallizes into a bcc crystal at low temperatures. This motivated the design of a binary version of it, which is presented in this paper. As, to our knowledge, the model is new, this paper is intended to be an introductory paper, focussing on the definition of the interactions and its mathematical properties, and on the structural properties of an equimolar mixture.

The paper is organized as follows: In section \ref{sec:def} the model is introduced. The next section \ref{sec:gen} details general results on thermodynamical observables: Equation of state, dependence of the mean energy on $T$ and polydispersity, chemical potentials and their relation with the radial distribution functions (rdf). We show also there that the unmixing of such a binary fluid is always prevented.
We consider finally the structural properties of the fluid in section \ref{sec:str} and show that both the partial rdfs and structure factors have particular splitting properties, reminiscent of additive mixtures of pair potential systems.
Several appendices give details on how the various properties announced in the main body of the paper can be derived.

\section{The polydisperse Voronoi fluid}
\label{sec:def}
The definition of the polydisperse Voronoi fluid begins with $N$ point particles, whose positions are denoted by $\bm r_i$ and velocity by $\bm v_i$ ($1\leq i\leq N$), evolving in a volume of dimension 3 with periodic boundary conditions (the generalization to a dimension $d\neq 3$ is straightforward). To each particle is associated a positive scalar, denoted by $R_i$ and termed ``natural radius'', which plays a role in the definition of interactions. We will see that this natural radius however does not define  a proper {\em excluded} volume.

\begin{figure}[h]
  \centerline{\begin{minipage}[t]{\textwidth}\input{sketch_Laguerre.pstex_t}\hspace{0.5cm}\raisebox{0.7cm}{\includegraphics[width=0.5\textwidth]{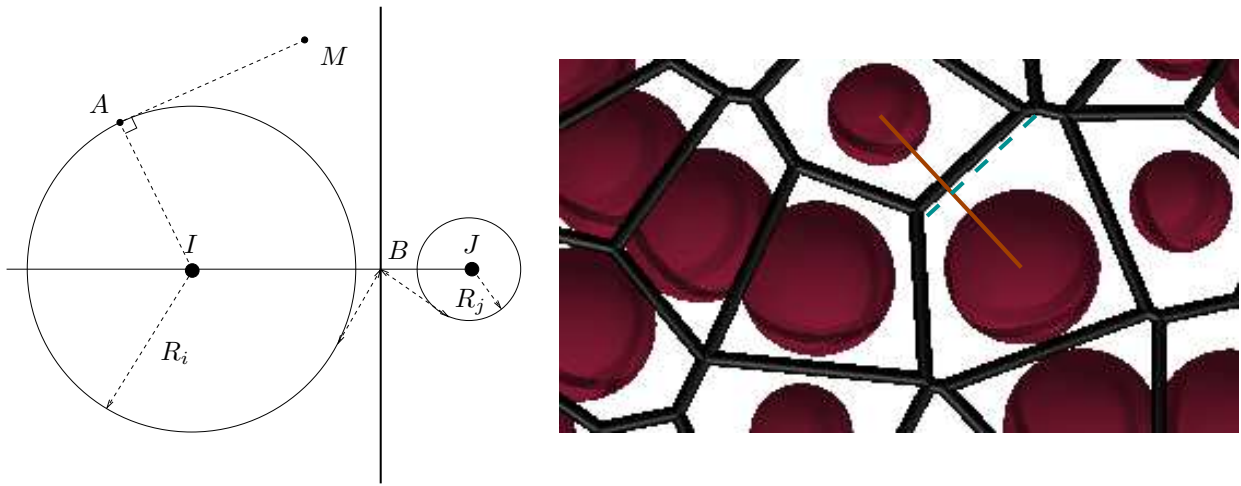}}
    \end{minipage}
  }
  \caption{Left : How a dividing plane is defined in the Voronoi-Laguerre tessellation. Two neighbouring particles $i$ and $j$, located at $I$ and $J$, and having natural radii $R_i$ and $R_j$ respectively, have Voronoi-Laguerre cells which share a common facet located in a plane normal in $B$ to $IJ$ (vertical bold line). The {\em powers} of $B$ (or any point of the plane) with respect to $i$ and $j$ are equal. Note that the dividing plane is shifted, in accordance with  intuition, toward the smaller particle. One has $IB=\rho_{ij}IJ$ with $\rho_{ij}=\demi[1+(R_i^2-R_j^2)/IJ^2]$. The figure  shows also  the distance $AM$ which represents the square root of the power of a point $M$ with respect to $I$. Right : Typical sketch of a 2D Voronoi-Laguerre tessellation. Dividing planes of neighbouring particles are unchanged with respect to the ordinary Voronoi, but shift toward the smallest particles otherwise (dashed blue : dividing line for an ordinary Voronoi tessellation).}
  \label{laguerre}
\end{figure}

To each configuration of the $N$ particles (weighed by their radius) one can associate a so-called Voronoi-Laguerre (or {\em radical}) tessellation : first, one defines a (pseudo) distance between the particle $i$ (located at $I$) and a point $M$ by the power of $M$ with respect to the sphere of center $I$ and radius $R_i$, namely $IM^2-R_i^2$. The geometrical interpretation of this distance is easy if $M$ is outside the sphere : in that case, the power is $AM^2$, where $A$ is the intersection point of the sphere and its tangent plane containing $M$ (see Fig. \ref{laguerre} left). If $M$ is inside the sphere, the power is negative, and its absolute value is the square of the radius of a circle drawed on the sphere with center $M$.
The cell of volume $v_i$ of the particle $i$ in the Voronoi-Laguerre tessellation is defined as the points of geometrical space whose power with respect to the particle $i$ are smaller than the power with respect to any other particle of the configuration. This corresponds to the definition of cells in the usual Voronoi tessellation with the only difference that the power of points with respect to the weighed particles replaces the usual distance to the particles. It is to note that the Voronoi-Laguerre tessellations of particles having all the same radius (a ``monodisperse'' Voronoi fluid) reduce to the ordinary Voronoi tessellation \cite{voronoiliquid}. Moreover, the natural radii of the particles have an effect on the dividing plane position only when two particles with different radii are neighbours  (see Fig. \ref{laguerre} right). This will have an interesting consequence for the stability of mixtures.

For the polydisperse Voronoi fluid, the interactions between the particles are defined via the potential energy
\begin{align}
  E_p(\bm r_1,\ldots,\bm r_N)&=\sum_{i=1}^N e_i=\frac{\ga}{2}\sum_{i=1}^N \int_{v_i} {\rm d}^3\bm r (r^2-R_i^2+R^2)\label{poten}
\end{align}
where in this expression $\bm r$ is a vector joining the particle $i$ to a  point spanning the cell $v_i$, and $R^2=N^{-1}\sum_{i} R_i^2$ is the mean squared radius of the particles (the term $\propto R^2$ in the energy is put to ensure an independence of the energy with respect to $R_i^2$ in the monodisperse limit, but,  being proportional to the total volume $V$, it does not affect the dynamics). We show in the appendix  \ref{app:forces} that the force $\bm{F}_i=-\bm\nab_i E_p$ acting on the particle $i$ is proportional to the so-called geometrical polarization $\bm \tau_i$ \cite{voronoiliquid}, namely a vector joining the particle to the barycenter of its Laguerre-Voronoi cell :
\begin{align}
  \bm F_i=\ga \bm\tau_i=\ga \int_{v_i} {\rm d}^3 \bm r\  \bm r\label{forces}
\end{align}
The so defined interactions  are local, translationally and rotationally invariant. At variance with usual 2-body potentials, they are intrinsically many-body. If one restricts oneself to binary mixtures only, with $N_1$ ``large'' particles of radius $R_1$ and $N_2=N-N_1$ ``small'' particles of radius $R_2<R_1$, it can be seen   that the potential energy  of a configuration (and hence the forces) depends on the intrinsic radii via the length $\sqrt{R_1^2-R_2^2}$ only:  This comes by inspection of the formula \myref{rhoij} and by noticing that $R^2-R_1^2=-(R_1^2-R_2^2)(N_2/N)$ (and a similar formula for $R^2-R_2^2$). Henceforward, we will make use of the dimensionless ``polydispersity parameter'' $\xi=(R_1^2-R_2^2)/v^{2/3}$, where $v=V/N$ is the average volume per particle. $\xi=0$ corresponds to the monodisperse fluid. It is a simplifying feature of this system that the polydispersity of a binary system is encoded by only one extra parameter.

\medskip

Throughout this paper, we consider a binary Voronoi liquid made of $\al_1=50$\% (we use the notation $\al_j=N_j/N$) of large particles with an polydispersity parameter $\xi=0.14$ at the density $v^{-1}=N/V=1$ . The dimensional factor $\ga$ is taken equal to 1000, so that the relevant temperatures are around $T\sim 1$ (we consider $k_B=1$ throughout the paper). The simulations were done using the LAMMPS code \cite{Plimpton}, modified  to compute the polarisations with the Voro++ library \cite{voro++}. We considered systems with $N=1000$ (sometimes $N=8000$) particles in cubic boxes with periodic boundary conditions. The systems were thermalized using a Nos\'e-Hoover thermostat, which samples the canonical $(N,V,T)$ ensemble. The temperatures considered range from $T=0.85$ to $T=2.00$, an interval which probes the moderately supercooled regime \cite{CavagnaReport} for our system (based on preliminary measurements, one expects a dynamical Arrhenius regime for $T\in [1, 2]$, an onset temperature around $T=1$, and a critical temperature of mode-coupling theory $T_c\simeq 0.8$ (based on the disappearance of the negative directions associated to the neighbouring saddles in the phase space).

\medskip

The parameter $\xi$, which  alone determines the degree of polydispersity, must be chosen with care. Actually, too large values of $\xi$ (or too large temperatures for a given $\xi$)  would lead to   physically unsound situations: Equation \myref{rhoij} shows that
for two neighbouring particles $i$ and $j$ with different radii and $r_{ij}^2/v^{2/3}<\xi$, the smallest particle is actually outside its Voronoi cell, {\em if any} :  Its volume could even vanish if all its neighbours were large and close by \cite{SpatialTessellations}. We are interested in supercooled systems, where steric hindrance dominates the relaxation phenomena. As a result, we  chose $\xi<1$ and moderate temperatures, so that the particles will always reside within their Voronoi cell, and the natural radii will de facto play the role of characteristic lengths of the effective soft repulsive core  (this latter point is however  rather subtle and indirect, since the cell boundary of two adjacent particles with the same radius is independent of that radius --- see paragraph \ref{sectiongder}). In the opposite limit of vanishing values of $\xi$, we would recover the monodisperse Voronoi fluid. Therefore, too small values of $\xi$ (or, equivalently,  too low densities) lessen the effect of the polydispersity and have to be avoided to suppress  crystallisation. In other words, Voronoi particles must be ``in contact'', but not too much. This point can be checked on the mixed radial distribution function $g_{12}(r)$ (see section \ref{sectiongder}), which has to be zero for distances $r<|R_1^2-R_2^2|^{1/2}$, and at the same time must have a first structural peak  at a distance $r^*_{12}$ not too large with respect to $|R_1^2-R_2^2|^{1/2}$. Our choice, $\xi=0.14$ and $v=1$, corresponds to $|R_1^2-R_2^2|^{1/2}=0.375$ and $r_{12}^*\simeq 1.12$ and fulfills this double requirement.

\section{General results on thermodynamics}
\label{sec:gen}

\subsection{Scaling properties and state equation}

In \cite{voronoiliquid}, we showed that the monodisperse ($\xi=0$) Voronoi fluid has unusual scaling properties, namely that the excess free energy (with respect to the ideal gas) has  the scaling form $F_e(T,V,N)=NT\phi[x=T/(\ga v^{5/3})]$, where $T$ stands for $k_BT$. A similar relation also holds for the {\em binary} Voronoi fluid, namely
\begin{align}
  F_e(T,V,N_1,N,R_1^2-R_2^2)&=NT\phi(x,\xi,\al_1)\label{scal}
\end{align}
where $x=T/(\ga v^{5/3})$, $\xi=(R_1^2-R_2^2)/v^{2/3}$ and $\al_1=N_1/N$ ({\em excess} means in excess with respect to the ideal mixing of ideal gases). On deriving with respect to $T$, one gets the excess entropy $S_e=-N[\phi+x(\pa\phi/\pa x)_{\xi,\al_1}]$. The  relation  $F_e=U_e-TS_e$ yields the excess internal energy, which is also the mean potential energy :
\begin{align}
  U_e&\equiv \lan E_p\ran=-NTx\dpar{\phi}{x}{\xi,\al_1},\label{ue}
\end{align}
 a relation one had already for the monodisperse fluid \cite{voronoiliquid}. Whereas a direct proportionality exists for the monodisperse fluid between $U_e=\lan E_p\ran$ and the excess pressure $P_e=-(\pa F_e)/(\pa V)=P-T/v$, this is no longer the case for the present binary fluid. One has instead (see Appendix \ref{app:pressure})
\begin{align}
  P_ev=-\frac{5}{3}\frac{\lan E_p\ran}{N}+\frac{2}{3}T\xi\dpar{\phi}{\xi}{x,\al_1}=-\frac{5}{3}\frac{\lan E_p^{(m)}\ran}{N}-\frac{\lan E_p^{(a)}\ran}{N},\label{Pev}
\end{align}
where $E_p^{(m)}$ (the main part) and $E_p^{(a)}$ (the additional part, so named because it is zero in the monodisperse limit) are defined by
\begin{align}
  E_p^{(m)}&\equiv\frac{\ga}{2}\sum_i\int_{v_i} {\rm d}^3\bm{r}\ r^2\\
  E_p^{(a)}&\equiv E_p-E_p^{(m)}=\frac{\ga}{2}\sum_i v_i(R^2-R_i^2)\end{align}

\begin{figure}[h]
  \centerline{\includegraphics[width=0.5\textwidth]{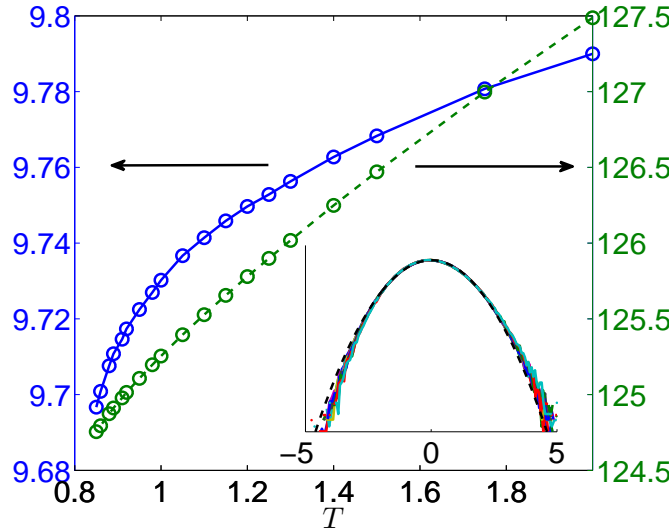}}
  \caption{Temperature dependence of : (Main) $\lan E_p^{(m)}\ran/N$ (dashed green) and $-\lan E_p^{(a)}\ran/N$ (solid blue), (Inset) Normalized distribution of particle energies $e_i$ for the large particles (see text for details). { All} temperatures collapse approximately on the Gaussian $y=\exp(-x^2/2)/\sqrt{2\pi}$ (black dotted). Note the small positive skewness.}
  \label{nrj}
\end{figure}

The temperature dependence of both components (per particle) is plotted in Fig. \ref{nrj} (notice that $-\lan E_p^{(a)}\ran$ is plotted to deal only with  positive ordinates).
The main part $E^{(m)}$ deserves its name, since one can observe that it comprises $\sim 92\%$ of the total energy. Being quite close to the total potential energy it is an increasing function of temperature, to ensure the positivity of $c_v^e$, the excess part of the constant volume heat capacity \cite{callen,voronoiliquid}.

The additional part $E_p^{(a)}$ is negative on average and decreases with temperature. Its negativity is easily understood if one recasts its expression, using $R^2=\al_1 R_1^2+\al_2 R_2^2$, into $E_p^{(a)}=\frac{\ga}{2}(R_1^2-R_2^2)N_1(V/N-V_1/N_1)$, where $V_1$ is the cell volume of particles with radius $R_1$. As we show below, the large cells (of type ``1'') must be in average larger than those of type ``2'' and consequently their mean volume $v_1=\lan V_1\ran/N_1$ is larger than the mean volume per particle $v=V/N$, leading to $\lan E_p^{(a)}\ran <0$. The magnitude of $\lan E_p^{(a)}\ran$ corresponds to a variation of $\simeq 10$\% of $v_1$ with respect to $v$. It is interesting to note that on cooling the system, the cell volume disparity between particles with different radii tends to lessen, contrary to the intuition (for instance, at very high temperatures, the potential energy, and thus the radii, do not matter anymore \cite{voronoiliquid}, and $E_p^{(a)}$ should go to zero). This paradoxical trend remains however modest in amplitude. The inset of fig. \ref{nrj} shows the scaled distribution function of the local energy $e_i$, shifted and scaled by the second moment, namely $\sig_eP(e_i=e)$ as a function of $(e-\lan e_i\ran)/\sig_e$ with $\sig_e=\sqrt{\lan e_i^2\ran-\lan e_i\ran^2}$. This distribution is almost insensitive to the temperature, and quite well described by a Gaussian (despite a small positive skewness). The typical values of $\sig_e$ are $\sim$ 6 times larger than the standard deviation of the total potential energy divided by $\sqrt{N}$, indicating a marked anticorrelation of neighbouring particles. The local energy distribution  is closely linked to that of the individual Voronoi volumes $v_i$, formerly investigated in \cite{StarrSastryDouglasGlotzer} for a bead-spring polymer model. In contrast with the findings of \cite{StarrSastryDouglasGlotzer} (where log-normal distributions were found  for several models of fluids), the  Voronoi volumes of the individual particles (large and small particles considered separately) are in our system extremely well described by a Gaussian distribution (without discernible skewness in fact) for all temperatures, stressing the direct impact of the nature of interactions on the statistical properties of the Voronoi related features, and the originality of the Voronoi fluid in this respect.

\medskip

An interesting property of the energy is its dependence on the polydispersity parameter, cf. \myref{dixneuf} :
\begin{align}
  \left(\frac{\pa E_p}{\pa \xi}\right)_{\bm r,V,N,\al_1}&=\frac{1}{\xi}E_p^{(a)}=\frac{\ga v^{2/3}}{2}(V\al_1-V_1)\label{Eprime}
\end{align}
where the derivative is taken keeping the positions of the particles fixed.
From this equation,
we can  deduce an explicit expression for the second derivative :
  \begin{align}
  \left(\frac{\pa^2 E_p}{\pa \xi^2}\right)_{\bm r,V,N,\al_1}&=-\frac{\ga v^{4/3}}{4}{\sum_{\lan i,j\ran}}'\frac{S_{ij}}{r_{ij}}\label{theprec}
\end{align}
where in the last expression, the primed sum must be restricted to neighbouring pairs of particles $(i,j)$ such that $R_i=R_1$ and $R_j=R_2$. One sees from the last two equations that for each configuration $E_p(\xi)$ is a concave function, whose Legendre transform $\{G(\la)=E_p(\xi)-\la\xi\ ;  \la=\pa E_p/\pa\xi\}$ is precisely $E_p^{(m)}$. This interpretation shows that $\xi$ is an intensive parameter conjugated to (a linear function of) $V_1$. The concavity of $E_p(\xi)$ has an important consequence : By the relation $\pa^2 F_e/\pa \xi^2=\lan \pa^2 E_p/\pa \xi^2\ran-\be \lan \de(\pa E_p/\pa \xi)^2\ran$ (with $\de A\equiv A-\lan A\ran$), one has that $F_e(\xi)$ is also concave. The fact that $\pa F_e/\pa \xi=\lan \pa E_p/\pa \xi\ran$ is zero for $\xi=0$ implies that $F_e$ is a decreasing function of $\xi$ for $\xi>0$. As a corollary of this property, one has always $\lan V_1\ran \geq V\al_1$ : The {\em individual} volumes of the particles with the larger natural radius are on average  larger than those of the smaller\footnote{Actually, one could even go one step further, since it can be shown that the facets $S_{ij}$ are second order polynomials of $\xi$. The expressions become howevcer involved.}.

\bigskip

The expressions \myref{Eprime} and \myref{theprec} are useful to investigate the polydisperse states for small deviations from  monodisperse equilibria, using a small $\xi$ expansion. We will for instance use them in section \ref{sectiongder} to understand the regular splitting of the different radial distribution functions at moderate polydispersity. As regards the thermodynamics,  the excess free energy (and all derived thermodynamic quantities) are expanded up to the second order in  $\xi$ :
\begin{align}
  F_e(\xi)/N&=F_e(\xi=0)/N-\frac{\ga v^{4/3} \al_1\al_2\xi^2}{8}\left(\lan \ell\ran_0+\frac{\ga}{T}\lan \De v^2\ran_0           \right) +o(\xi^2)\label{expa}\\
   \ell&\equiv \frac{1}{N} \sum_{i,j/\lan i,j\ran} \frac{S_{ij}}{r_{ij}}\\
    \De v^2&\equiv \frac{1}{N}\sum_{i}[v_i^2-v^2]
\end{align}
  where $\lan\ldots\ran_0$ is a canonical average over the monodisperse system ($\xi=0$), and $\de X\equiv X-\lan X\ran$.
  \begin{figure}[h]
\centerline{\includegraphics[width=0.5\textwidth]{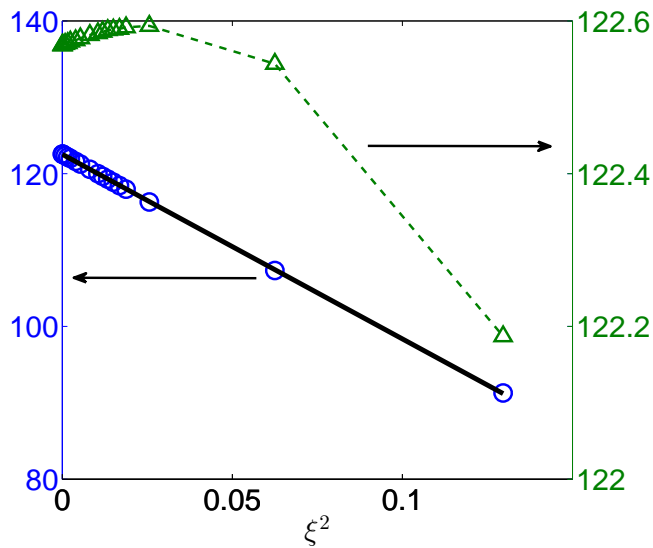}}
\caption{Blue circles : Mean potential energy per particle as a function of ${\xi}^2$. The temperature is $T=2$. The black line is the best linear fit with slope -242. Green triangles : $\lan E_p-\demi E_p^{(a)}\ran$ as a function of ${\xi}^2$ for $T=2$.}
\label{toutdroit}
  \end{figure}

  The expansion \myref{expa} is interesting in several ways. First, it shows that the polydispersity has only a modest impact on the average energy for low values of $\xi$. In fig. \ref{toutdroit}, we plotted $\lan E_p\ran/N$ against $\xi^2$ (blue circles), together with a linear fit. The quality of the fit is striking until values of $\xi$ as high as $0.4$.  Second, in this limit, the dependence with respect to the composition is explicitly given by the factor $\al_1\al_2$ multiplying $\xi^2$, and shows a symmetry under an identity exchange  between particles of different types (due to the fact that this exchange is exactly equivalent to make $\xi\rightarrow-\xi$). Third, \myref{expa} shows that the free energy dependence on polydispersity is quantitatively determined by two  terms, coming from the second derivative of $E_p$ with respect to $\xi$ and fluctuations of the first derivative (term $\propto \lan \De v^2\ran$). Note that the limit $T\rightarrow0$ is pathological within this formula, since it predicts a $T^{-2}$ negative divergence of the excess entropy. However, at the lowest temperature tested ($T=0.85$), the two terms are of the same order of magnitude ($\lan \ell\ran=4.8$ and $\frac{\ga}{T}\lan\De v^2\ran=2.6$).

  A final comment concerns the triangles of fig. \ref{toutdroit}. It shows that, at the temperature $T=2.0$, $\lan E_p-E_p^{(a)}/2\ran$ is almost independent of $\xi$ in the range considered.  Clearly, this is not an exact result, but it can be understood if one assumes $\pa ^2 E_p/\pa \xi^2$ (cf. \myref{theprec}) almost independent of $\xi$ for the typical configurations (actually this quantity is a second order polynomial in $\xi$). We checked numerically that this assumption is correct, by directly monitoring the $\xi$ dependence of $V_1$ of representative configurations of the fluid. The constancy of \myref{theprec} with respect to $\xi$ yields immediately that of $\lan E_p-E_p^{(a)}/2\ran$.


  
\subsection{Strong mixing}

An interesting corollary of the formula \myref{Eprime}  is that  the liquid will remain mixed whatever its composition. This property comes from the observation that a demixed polydisperse Voronoi fluid is, apart from the surface boundary between the two phases (and hence negligible at the thermodynamic limit), fully equivalent to the {\em monodisperse} fluid! This is because the Voronoi cell of a particle with a given radius, surrounded by identical particles, is independent of the value of that radius. As a result, one has at the thermodynamic limit, assuming $R_1>R_2$:
\begin{align}
  F_e^{\rm (mixed)}-F_e^{\rm (unmixed)}&=F_e(\xi)-F_e(\xi=0)<0\label{gloubi}
\end{align}
as shown in  the preceding paragraph.
As a result, the excess free energy of the unmixed state is always greater than that of the mixed state, and therefore  cannot compensate the unmixing entropy cost which would be present in the ideal term.

Very close to the monodisperse state, the relation equivalent to \myref{gloubi} is
\begin{align}
  \left.\frac{\pa^2F_e}{\pa \xi^2}\right|_{\xi=0}&=\frac{\pa}{\pa \xi}\left\lan\frac{\pa E_p}{\pa \xi}\right\ran
                                                 =-\frac{N\ga v^{4/3}\al_1\al_2}{4}\left\{ \lan\ell\ran_0+\frac\ga{T} \lan \De v^2\ran_0\right\}<0
\end{align}


\subsection{Chemical potentials}

A third derivative of the free energy gives access to the chemical potential : the excess chemical potential of species $1$ is  given by
\begin{align}
  \mu_{e,1}&=g_e+T\dpar{\phi}{\al_1}{x,\xi}\al_2=\mu_{e,2}+T\dpar{\phi}{\al_1}{x,\xi}
\end{align}
where $g_e=T\phi+P_ev$ is the Gibbs enthalpy per particle. As for the monodisperse system \cite{voronoiliquid}, a kind of ``zero-separation theorem'' applies here, namely a configuration with $N_1-1$ particles of type 1 can be obtained by superimposing two such particles on top of each other. The specificity of the present case comes from the three possibilities one has for this ``Gedankenexperiment'' : one can superimpose two type 1 particles, two type 2 particles, or a type 1 particle on a type 2 one. From the first two possibilities, one gets the relation \cite{voronoiliquid}
\begin{align}
  \mu_j&=k_BT\ln g_{jj}(r=0)\ \text{for } j\in\{1,2\}
\end{align}
where $g_{jj}(r)$ is the radial distribution function of particles $j$ only. Recalling that we assume $R_1>R_2$, the third option yields $\mu_2=k_BT\ln g_{12}(r=0)$. This asymmetry between the two types of particles comes from the fact that when superposing two particles with different Voronoi-Laguerre radii, the tessellation is entirely given by the particle with the largest radius, the other one having an ``empty'' Voronoi cell. Unfortunately, these relations are of little use at low temperature where an effective repulsion between the particles prevents them from coming close to each other, and makes the vicinity of $r=0$ statistically irrelevant for the $g_{ij}(r)$.

\subsection{Constant volume heat capacity and isothermal compressibility}

\begin{figure}[h]
  \centering
  \includegraphics[width=0.5\textwidth]{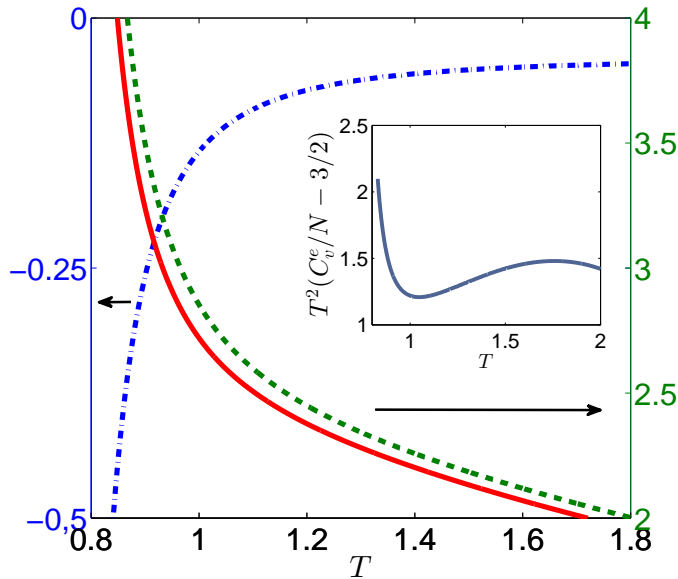}
  \caption{Temperature dependence of $C_v^e/N$ (red solid) and of the two components $C_v^{(x)}/N=N^{-1}d\lan E_p^{(x)}\ran/dT$ of the excess constant volume heat capacity $C_v^e/N$: Dashed green, $x=m$; Dash-dotted blue : $x=a$. Inset : test of the ``quadratic configurational entropy hypothesis'' prediction for $C_v^{(e)}$. See text for details.}
  \label{figcv}
\end{figure}

In terms of the scaling function $\phi$ (eq. \myref{scal}), one has from \myref{ue} for the excess constant volume heat capacity $C_v^e$ :
\begin{align}
  C^e_v/N&=-\pa_x[x^2\pa_x\phi]
\end{align}
As the energy has the additive decomposition $E_p=E_p^{(m)}+E_p^{(a)}$, one can  write $C_v^e=C_v^{(m)}+C_v^{(a)}$, but this decomposition loses track of the deep connection which  holds between the excess heat capacity and energy fluctuations: Firstly $C_v^{(a)}$ is negative, secondly $C_v^{(m)}$ is not proportional to $\lan [\de E_p^{(m)}]^2\ran$. The correct relations linking these partial capacities and fluctuations become $T^2C_v^{(x)}=\lan [\de E_p^{(x)}]^2\ran+\lan\de E_p^{(m)}\de E_p^{(a)}\ran$ (with $x\in (a,m)$).

In fig. \ref{figcv} are plotted the two components $C_v^{(m)}$ and $C_v^{(a)}$ of the heat capacity as a function of $T$. Due to the fact that $E_p^{(a)}$ is proportional to the (small) polydispersity parameter $\xi$, the ``additional'' contribution $C_v^{(a)}$ gives only a minor negative correction to the overall positive result. Both curves  show a marked crossover between a high temperature regime with mild variations of the specific heats and a low temperature regime where a steep increase of the $C_v^e$ with decreasing $T$ can be observed. This steep increase can be understood within the framework of the potential energy landscape (PEL) decomposition, wherein the configurational space is partitioned into basins of attraction of the different potential energy minima (so-called inherent structures (IS)). At low temperatures, a decoupling of the thermodynamics between the IS and the vibrational degrees of freedom within each IS basin is assumed, harmonic vibrations being preponderant at very low $T$. This theoretical argument leads to the prediction
\begin{align}
  C_v^e&=\frac{\pa\lan E_{IS}\ran}{\pa T}+\frac{3N}{2}
\end{align}
for the excess specific heat, where $\lan E_{IS}\ran$ is the mean value of the IS energy at temperature $T$. In many cases \cite{HeuerReview}, the configurational entropy is well described in the low $T$ regime by a quadratic formula, which implies that $\lan E_{IS}\ran$ is linear in $1/T$: $\lan E_{IS}\ran=C^t-N\sig^2/T$. This would imply thus $C_{v}^e/N=\sig^2/T^2+3/2$. This hypothesis is tested in the inset of fig. \ref{figcv}: Clearly, the agreement is not good. We postpone an analysis of this problem to a future publication, but preliminary results indicate that for our system, the configurational entropy is very satisfactorily described by a quadratic function (with  $\sig^2\simeq 1$), whereas anharmonic terms seem to remain quite important even at low temperature.

\medskip

Let us now consider the isothermal compressibility $\chi_T=-V^{-1}(\pa V/\pa P)_T$, or, more conveniently, the (zero-frequency) bulk modulus $\chi_T^{-1}$. Contrary to the compressibility, the bulk modulus splits into an ideal part  $T/v$ and excess part $-V(\pa P_e/\pa V)_T$. For the binary Voronoi liquid, this part can be expressed as
\begin{align}
  (\chi_T^{-1})^{(e)}&=-\frac{2}{3}P_e-\left(\frac{5}{3}\right)^2\frac{TC_v^{(e)}}{V}+\frac{\ga}{3}(R_1^2-R_2^2)\left[\dpar{ V_1}{ V}{N,T}-\al_1-\frac{5T}{3V}\dpar{V_1}{T}{N,V}\right]
\end{align}
Quantitatively, the three terms of the right hand side of the last equation are quite different in the low temperature regime we are interested in. The last term is $O(\xi^2)$, strictly zero for the monodisperse fluid, therefore negligible in our system with a rather small polydispersity. The second term  is quantitatively $\simeq  -9$ for $T=0.85$. The first term is $\simeq 150$, and thus is responsible for a low compressibility ($\chi_T T/v\simeq 6\cdot 10^{-3}$) of the Voronoi fluid: This value for $\chi_T T/v$ is typically more than three times smaller than that of the simple fluids close to the triple point \cite{Hansen,Verlet165}.

\subsection{Stress tensor}

The stress tensor of this model could be a priori  a complicated observable, due to the fact that the forces are not pairwise. In such cases, a systematic procedure \cite{admaltadmor} has been devised to get a {\it bona fide} microscopic stress tensor. This procedure would be rather involved for our model, but fortunately, a direct recognition of the stress tensor is possible.

To begin with, the very definition of the forces, namely vectors proportional to the local geometrical polarizations, together with the polyhedral nature of the Voronoi cells, allow a natural decomposition into a sum of pseudo-pair forces directed along the vectors joining two neighbouring particles. Indeed, a variant of the Green-Ostrogradski (GO) theorem yields
\begin{align}
  \bm F_i&=\ga\int_{v_i}{\rm d^3}\bm r\ \bm r=\frac{\ga}{2}\oint_{\pa v_i}{\rm d}^2\bm S\cdot [r^2-R_i^2+R^2]=\sum_{j/\lan j,i\ran}\underbrace{\frac\ga{2} \int_{S_{ij}}{\rm d}^2\bm S\cdot [r^2-R_i^2+R^2]}_{\equiv\bm F_{j\rightarrow i}}\label{Fji}
\end{align}
This decomposition of the total force $\bm F_i$  as a sum of pair terms $\bm F_{j\rightarrow i}$ verifies the properties expected from such a decomposition : one has $\bm F_{j\rightarrow i}$ parallel to $\bm r_{ij}$ and $\bm F_{j\rightarrow i}=-\bm F_{i\rightarrow j}$ by a simple verification (the term proportional to $R^2-R_i^2$, absent in the monodisperse system \cite{papersound}, is required here to ensure the latter antisymmetry). In other words, this pair force decomposition complies with the so-called strong third Newton law (the law of action-reaction with $\bm F_{j\rightarrow i}$ parallel to $\bm r_{ij}$).  For a monodisperse fluid, this effective pair force accounts for an {\em attractive} interaction between individual particles, even at very short distances. For a small polydispersity $\xi$, one expects this property to hold as well. We can then envision the Voronoi liquid as being a fluid {\em constantly under tension} (thus with a strong negative pressure), the cavitation being prevented by an effective infinite surface tension (the work associated with the creation of an empty bubble of radius $R$ is $\sim \ga R^{5}$, and that bubble is never mechanically equilibrated).

\medskip

With the decomposition \myref{Fji}, it is tempting to define  the (hydrodynamic) $zz$ component of the excess stress tensor by
\begin{align}
  \sig_{zz}^{\rm e}&=\demi \sum_{i}\sum_{j,j\neq i} \tilde{z}_{ij}F_{i\rightarrow j,z}\label{classicalstress}
\end{align}
where $\tilde{z}_{ij}=\bm r_{ij}\cdot \bm e_z-L_z[(\bm r_{ij}\cdot \bm e_z)/L_z]$ is the $z$ component of the vector $\bm r_{ij}$ compliant with the minimum image convention ($[x]$ holds here for the integer nearest to $x$), and $F_{i\rightarrow j,z}=\bm F_{i\rightarrow j}\cdot \bm e_z$. We showed  in a preceding work \cite{papersound} that the decomposition of the force to a sum of partial forces obeying the strong third Newton law {\em does not guarantee} that \myref{classicalstress} is a correct expression for the $zz$ component of the stress tensor. Fortunately, this is however true here, as shown in Appendix \ref{app:stress}.

\subsection{Hessian of the potential energy function}

When the temperature of a glass-forming liquid is lowered to the mode-coupling temperature $T_c$ or less, the relaxation dynamics becomes primarily activated, with the system oscillating for substantial period of times around a minimum of potential energy. This has been shown for instance in \cite{BroderixSaddles,AngelaniSaddles}, where the typical saddles close to which the system is located in course of time lose their negative directions for $T\leq T_c$. As a result, the Hessian of the potential energy is an important observable  to determine in this context. Fortunately, this is quite easy for the Voronoi fluid, see formula \myref{hesstens}. As a result, if $\bm r_i^0$ are the positions of a critical point of $E_p(\{\bm r_i\}_{i\in\{1..N\}})$ (that is, an equilibrium point, stable or unstable (saddle)), we can write, for small displacements $\de\bm r_i$ from the critical point (we define also $\de \bm r_{ij}=\de \bm r_j-\de\bm r_i$)
\begin{align}
  E_p(\bm r_i^0+\de\bm r_i)-  E_p(\bm r_i^0)&=-\frac{\ga}{2} \sum_{i<j} \frac{1}{r_{ij}}\int_{S_{ij}}{\rm d}S(\de \bm r_{ij}\cdot\bm r)(\de \bm r_{ij}\cdot\bm r')\\
  &=\frac{\ga}{2}\sum_{i<j} S_{ij}\frac{\rho_{ij}\rho_{ji}}{r_{ij}}(\de \bm r_{ij}\cdot\bm r_{ij})^2-\frac{\ga}{2}\sum_{i<j}\frac{1}{r_{ij}}\int_{S_{ij}}{\rm d}S(\de \bm r_{ij}\cdot\bm r_\parallel)^2\label{hess1}
\end{align}
where $\bm r_\parallel$ is the component of the vector $\bm r$ spanning the facet $S_{ij}$, parallel to that facet, and $\bm r'=\bm r_{ji}+\bm r$. From this formula, it is  clear that the displacement of neighbouring pairs parallel to their axis enhances systematically the energy, whereas those parallel to their common face decreases the energy. For a system with interactions described by a pair potential $\phi(r)$, a similar formula exists:
\begin{align}
  E_p(\bm r_i^0+\de\bm r_i)-  E_p(\bm r_i^0)&=\demi \sum_{i<j}\phi''(r_{ij})(\de \bm{\tilde{r}}_{ij})^2+\demi\sum_{i<j} \frac{\phi'(r_{ij})}{r_{ij}}(\de \bm{\hat{r}}_{ij})^2\label{hess2}
\end{align}
with $\de\bm{\tilde{r}}_{ij}$ (resp. $\de\bm{\hat{r}}_{ij}$) is the component of $\de\bm r_{ij}$ parallel (resp. normal) to $\bm r_{ij}$. The  terms in the right hand side of this expression are successively positive and negative for a repulsive potential. Therefore the structuration of the response of the potential energy of the Voronoi fluid near an equilibrium point is not very different from that of a repulsive pair potential : the elongational component of $\bm r_{ij}$ is associated with a positive increment in energy, the shearing component to a negative one. Beyond the expressions which are obviously different, one can notice that regarding the perpendicular (shearing) component, the pair potential is only sensitive to the modulus of that component, whereas the Voronoi fluid is also sensitive to its direction, ultimately a consequence of the multibody nature of the potential.

Finally, a last comment concerns the quantitative balance between the two terms of the right hand sides of \myref{hess1} and \myref{hess2}. If $\phi(r)\propto r^{-n}$, the first term of \myref{hess2} is typically $n+1$ times larger than the second, provided $\de r_{ij,\parallel}\sim\de r_{ij,\perp}$. For the Voronoi liquid, a rough estimation using the arguments given at the end of Appendix \myref{app:stress} shows that the first term is roughly 16 times larger than the second (with the same proviso as before). Compared to the pair potential case, it would correspond to a stiff pair potential fluid with $\phi(r)\sim r^{-15}$ and an overlapping density. This is in line with the very low compressibility observed.

\section{Structural properties}
\label{sec:str}

\subsection{Radial distribution functions}\label{sectiongder}

\begin{figure}[h]
  \centerline{\includegraphics[width=0.5\textwidth]{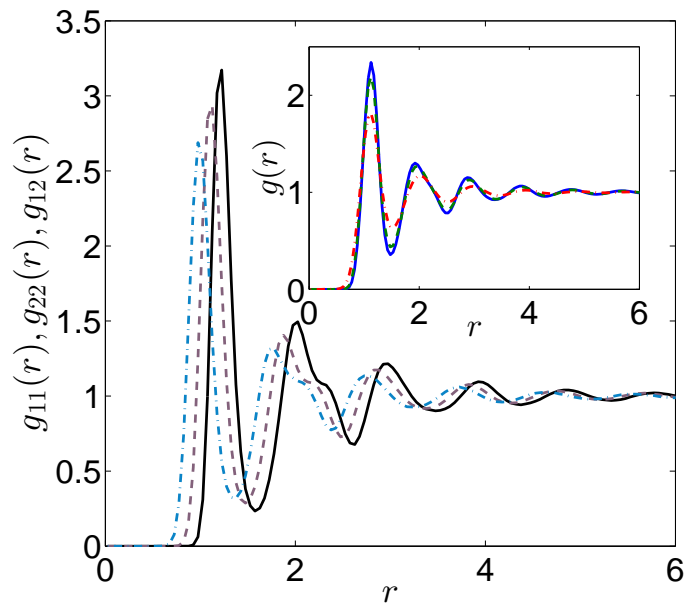}}
  \caption{Radial distribution functions of the Voronoi fluid. Main : for $T=0.85$ and $N=8000$, $g_{11}(r)$ (solid black), $g_{12}(r)=g_{21}(r)$ (dashed purple) and $g_{22}(r)$ (dash-dotted light blue). Inset : $g(r)$ for temperatures $T=0.85$ (solid blue), $T=1.00$ (dashed green) and $T=2.00$ (dash-dotted red).}
  \label{gder}
\end{figure}

The simplest  characterization of the microscopic structure is provided by the radial distribution function (rdf) $g(r)=(v/N)\sum_{i\neq j}\lan \de(\bm r-\bm r_{ij})\ran$, which can be split into $g(r)=\al_1^2g_{11}(r)+(1-\al_1)^2g_{22}(r)+2\al_1(1-\al_1)g_{12}(r)$ with $g_{ab}(r)=V/(N_a N_b)\sum'_{i\neq j} \lan \de(\bm r-\bm r_{ij})\ran$ where the primed sum is restricted to particles $i$ being of type $a$ and particles $j$ being of type $b$. This defines the partial rdf $g_{ab}(r)$, which are plotted in fig. \ref{gder} for the temperature $T=0.85$. In the inset, the total $g(r)$ is shown for the temperatures $T=0.85,1.00,2.00$.
The inset shows a usual temperature dependence of the $g(r)$, that is, a progressive structuration of $g(r)$ on cooling, with a growth of the first peak and a small shoulder at the second peak, characteristic of the dense random packing \cite{Wahnstrom}. The partial rdfs are slightly shifted with respect to each other, which reflects both the slight degree of polydispersity, and the high degree of mixing. Generally speaking, the salient features of the rdfs are more pronounced in the partial rdf than in the total one, because the average nature of the $g(r)$ and the dephasing causes a smoothing of their features: One sees for instance that the first peak's height is higher for the three partial rdfs than for $g(r)$, and the shoulder of the second peak is also more pronounced in each partial rdf.

The analysis of the shifts of the $g_{ab}(r)$ with respect to each other illustrates the collective nature of the thermodynamic equilibrium: We have already stressed the fact that the ``direct'' interaction of two neighbouring {\em identical} particles, that is, roughly speaking, the position of the dividing facet common to their Voronoi cells, is insensitive to the natural radius of these particles. As a result, the fact that two neighbouring large particles are typically farther apart than two neighbouring small particles (see the relative positions of the first peaks of $g_{11}(r)$ (large particles) and $g_{22}(r)$ (small particles) in fig. \ref{gder}) results from an overall balance of influences over the whole cell: If one considers for instance a large particle, the neighbouring small particles locally enlarge the Voronoi cell with respect to what it would be in an environment of large particles only. The pressure equilibration thus displaces the mean position of the large particle within its cell with respect to a neighbourhood of large particles, away from the large particles, closer to the small ones. A corresponding argument explains the contraction of the typical neighbouring small particles. As a result of this pressure balance, it is worth noting that the loci of the maxima of  $g_{ab}(r)$ are remarkably temperature independent in the temperature range probed ($T\in [0.85,2]$): $g_{11},g_{12}$ and $g_{22}$ are maximum at $1.225v^{1/3}$, $1.125v^{1/3}$ and $0.975v^{1/3}$, respectively. If one takes the first and third of these values as the effective diameters of particles $1$ and $2$, one gets an {\em a posteriori} aspect factor $\xi_{\rm post}=(0.371)^2$, not far from the {\em a priori} value $\xi=(0.375)^2$. These effective diameters are apparently additive, in the sense that the maximum position of $g_{12}$  is equal to  the average of the positions of the maxima of $g_{11}$ and $g_{22}$ (within 2\%). We {\em prove} this result in Appendix \ref{app:rdfmax}, in the limit of small polydispersities ($\xi \ll 1$). It is shown there also that it is not a consequence of the special ratio $\al_1=\al_2=1/2$  considered in this work, and that this property would hold true whatever the value of $\al_1$. Moreover, this ``equidistribution'' of the first maximum of the $g_{ab}$ extends actually to {\em all extrema}, and one can even show that this remarkable property holds also for the extrema of the structure factors $S_{ab}(k)$.

\subsection{Structure factors}

\begin{figure}[h]
  \centerline{\includegraphics[width=0.5\textwidth]{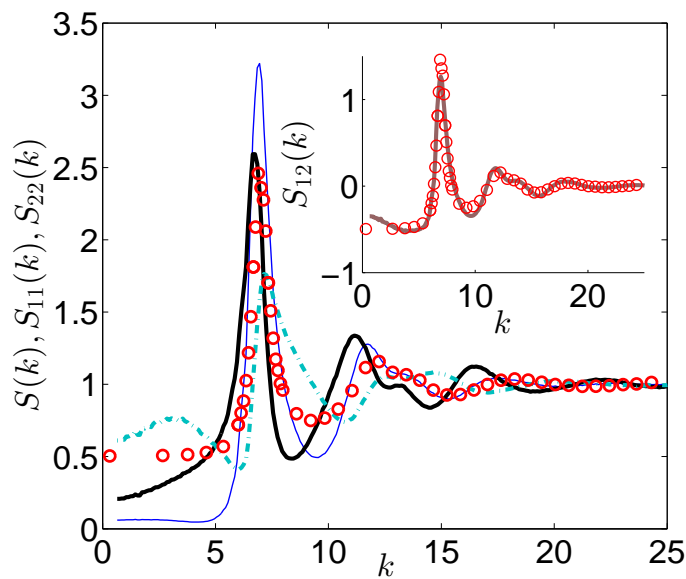}}
  \caption{Static structure factors of the Voronoi fluid. Main : for $T=0.85$ and $N=8000$, $S_{11}(k)$ (solid black),  $S_{22}(k)$ (dash-dotted light blue), $S(k)$ (thin solid ultramarine). The red circles show the ``random Ansatz'' $S^A_{11}(k)=\al_1S_{\rm mono}(k)+1-\al_1$ (see text for details). Inset : $S_{12}(k)$ for $T=0.85$ (solid purple), and the random Ansatz $S^A_{12}(k)=\sqrt{\al_1\al_2}(S_{\rm mono}(k)-1)$.}
  \label{sdek}
\end{figure}

The partial static structure factors (SFs) $S_{aa}(k)=N_a^{-1}\lan \rho_a(-\bm k)\rho_a(\bm k)\ran$ and $S_{ab}=S_{ba}=(N_aN_b)^{-1/2}\lan \rho_a(-\bm k)\rho_b(\bm k)\ran$, where $\rho_a=\sum_{j/R_j=R_a}\exp(i\bm k\cdot\bm r_j)$ is the Fourier component of the density field for the particle of type $a$ ($a=1,2$), are related to the total structure factor by $S(k)=\al_1S_{11}(k)+(1-\al_1)S_{22}(k)+2\sqrt{\al_1(1-\al_1)}S_{12}(k)$. In fig. \ref{sdek}
are shown the SFs for $T=0.85$. Interestingly, whereas the rdfs $g_{11}$ and $g_{22}$ seem  approximately related to each other by an approximate shift of their axes, the SFs $S_{11}$ and $S_{22}$, which convey the same overall information, are well different. More precisely, if the large particles have a partial SF reminiscent of a monodisperse compressible fluid (with however a atypically smooth crossover between  microscopic and macroscopic lengthscales), the small ones have a partial SF somewhat different, with a first peak closely associated to a dip, and a secondary mild maximum at $\sim k_0/2$ (where $k_0$ is the abscissa of the first peak of $S(k)$). If it is difficult to ascribe precisely these differences to a well-defined physical origin, it is however allowed to say that the organization of the large particles is less perturbed by the polydispersity, in the following sense: If, invoking the weak polydispersity, one assumes that the substructure of the large  particles can be envisioned as a substructure obtained from the cold monodisperse Voronoi fluid \cite{voronoiliquid} having a structure factor $S_{\rm mono}(k)$ (we took the structure factor for the coldest temperature of the metastable liquid, corresponding here to $T=1$), by picking {\em at random} $N\al_1$ particles within each configuration, one would have for $S_{aa}(k)$ the ``random Ansatz'' $S^A_{aa}(k)=\al_{a}S_{\rm mono}(k)+1-\al_a$ and $S^A_{12}(k)=\sqrt{\al_1(1-\al_1)}[S_{\rm mono}(k)-1]$ for $S_{12}(k)$ .

The random Ansatz $S_{11}^A$ and $S_{12}^A$ are shown  as red circles in fig. \ref{sdek}. One can see that the Ansatz  $S^A_{12}(k)$ is rather good, indicating that as far as cross correlations are considered, the picture of a random distribution of $\al_1N$ large particles among the total population of particles is sensible.

As regards   $S^A_{11}(k)$, it can be equally compared to the real curves for $S_{11}(k)$ and $S_{22}(k)$, since in our equimolar mixture  $S^A_{11}(k)=S^A_{22}(k)$. If one considers first the wavenumbers around or larger than the first structural peak, one sees that  $S^A_{11}(k)$ and $S_{11}(k)$ are semi-quantitatively similar (a shift of the minima and maxima can be observed, it corresponds to the already mentioned extension of the typical distances between type 1 particles), and certainly closer to each other than $S^A_{22}(k)=S^A_{11}(k)$ and $S_{22}(k)$. It is difficult to go beyond this qualitative discussion, which however shows that the distribution of the large particles is less perturbed by the polydispersity than the small particles.

The intermediate-to-small $k$ domain is different: Clearly, the ``random  Ansatz'' is incorrect for both  small and large particles. This shows that the repartition of large particles beyond the first coordination shell is not that of a random mixing. This can be traced back to the $g(r)$. The rdf of the monodisperse fluid at the coldest temperature $T=1$ is very close to the total one of the binary fluid for $T=0.85$, in particular the monodisperse $g(r)$ does not show the pronounced shoulder of the second peak of the partial rdfs, which evidences this additional, beyond-the-first-shell ordering.

\section{Conclusion}

In this paper, we have presented a new model of polydisperse fluid, based on a generalization of the so-called Voronoi fluid presented elsewhere \cite{voronoiliquid}. This fluid is defined via its potential energy, which requires the Voronoi-Laguerre tessellation of the configuration, an ``intrinsic radius'' associated to each particle modifying the tessellation and making the fluid effectively binary. We have presented a structural study of a 50:50 mixture of binary particles characterized by a polydispersity parameter $\xi=\sqrt{R_1^2-R_2^2}/v^{2/3}=0.14$, in the moderately low temperature range $T\in[0.85, 2.0]$ (a preliminary study indicates that the mode-coupling temperature of this model is around $T_c\sim 0.80$). The thermodynamics of this model has been studied via the equation of state, which shows a strong negative excess pressure. We recovered this result by finding an explicit expression for the microscopic stress tensor, which makes it clear that the fluid is constantly under tension (notice that no gas phase can coexist with the liquid \cite{voronoiliquid}).
By studying the temperature and $\xi$ dependences of the mean potential energy in the $NVT$ ensemble, we have been able to show rigorously that the convexity properties of the excess free energy with respect to $\xi$ imply a thermodynamic impossibility for the binary Voronoi fluid to phase separate. As regards the second order thermodynamic coefficients,  the computation of the excess heat capacity at constant volume has a particularly high value $C_v^e\simeq 3-4$ that might be related to a unusually large value of the variance of the configurational entropy at low temperature \cite{sciortinoJSTATMECH}. Furthermore, the compressibility is low, of order $10^{-2}-10^{-3}(v/T)$, the excess bulk modulus being mainly $\sim -P_e$. We finally studied the microscopic structural observables, namely the partial radial distribution functions and structure factors, to show the short-range liquid structure of the binary Voronoi liquid, with a unique additivity property at low values of $\xi$.

In future work, we will study the dynamics of this model, with a particular emphasis on the glass transition scenario. We have shown in \cite{papersound} that the dynamical properties of the Voronoi liquid could be quite different from that of ordinary liquids. It is interesting to see to what extent deviations with respect to the ``usual'' behaviour could also show up in the binary Voronoi liquid. For instance, the mode-coupling theory, in its usual presentation, does not a priori preclude fluids with manybody interactions from its scope. On the other hand, Zaccarelli et al. \cite{ZFGST} have shown that the MCT can be recovered from systematic Gaussian approximations on the Newtownian equations for the dynamics of density modes, {\em provided the interactions are pairwise additive}. As a result, testing the predictive efficiency of the MCT in  a  system with manybody interactions is an interesting point, since it could help narrowing the validity conditions of this theory. Another point which this paper raises is related to the quite high excess heat capacity. This feature of the model points to a large value of the variance $\sig^2$ of the configurational entropy \cite{sciortinoJSTATMECH}, a quantity which may be related to the dynamics through the Adams-Gibbs picture. A model where $\sig^2$ would show a marked contrast with respect to the usual models would be interesting in a comparative study.

\appendix

\section{Computation of the forces}\label{app:forces}

In this Appendix, we show how the formula \myref{forces} is derived from the definition of the potential energy \myref{poten}. First, one defines the local energy
\begin{align}
  e_i&=\int_{v_i}{\rm d}^3\bm r (r^2-R_i^2+R^2)
\end{align}
which can be recast
\begin{align}
  e_i&=\int {\rm d}\Om L_i(\bm n)^3\left(\frac 15 L_i(\bm n)^2+\frac 13(R^2-R_i^2) \right)\label{ei}
\end{align}
where ${\rm d}\Om$ is an elementary solid angle around a unit vector $\bm n$ and $L_i(\bm n)$ is the distance from $i$ to the boundary of its Voronoi cell in the direction $\bm n$ \cite{epjeVoronoi,makse}:
\begin{align}
  L_i(\bm n)&=\min_{j/ \bm n\cdot \bm r_{ij}>0} \frac{\rho_{ij}r_{ij}^2}{\bm n\cdot\bm r_{ij}}\label{rhoijmoinsun}\\
  \rho_{ij}&=\demi+\frac{R_i^2-R_j^2}{2r_{ij}^2}\label{rhoij}
\end{align}
with the definition $\bm r_{ij}=\bm r_j-\bm r_i$. Note that $\bm r_{ji}=-\bm r_{ij}$ but that $\rho_{ji}=1-\rho_{ij}$. A facet of the Voronoi-Laguerre tessellation shared by cells $i$ and $j$ is located by definition at a distance $\rho_{ij}r_{ij}$ (resp. $\rho_{ji}r_{ji}$) from particle $i$ (resp. $j$), and normal to $\bm r_{ij}$.

\begin{figure}[h]
  \centerline{\input{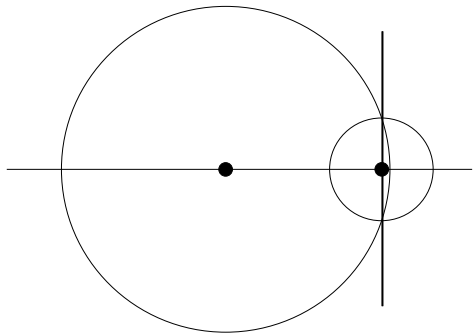}}
  \caption{Limiting case beyond which the right particle does not remain in its Voronoi cell.}
  \label{merdouille}
\end{figure}
These considerations assume that $\rho_{ij}>0$ and $\rho_{ji}>0$. This is no longer true  if the distance of particles with different radii is  $r_{ij}^2\leq R^2_1-R_2^2$. The limiting case ($r_{ij}^2=R_1^2-R_2^2$) is shown in Fig. \ref{merdouille}. This leads to complications in  the structure of the Voronoi cells (for instance, particles can be outside their own cell, or have no cell at all), and weakens considerably our interpretation of the natural radii as excluded volumes. As a result, we will consider only such situations where temperature and  density are not too high, such that for all neighbouring particles $i$ and $j$, we  {\em always} have $|R_i^2-R_j^2|<r_{ij}^2$.

If $i\neq j$, one has
\begin{align}
  \bm \nab_i e_j&=\int d\Om L_j(\bm n)^2\left(L_j(\bm n)^2+R^2-R_j^2\right)\nab_i L_j(\bm n)
\end{align}
This is zero if cells $i$ and $j$ do not have a common boundary, otherwise we have
\begin{align}
  L_j(\bm n)^{-1}\bm\nab_i L_j(\bm n)&=-\frac{\bm r_{ij}\rho_{ij}}{\rho_{ji}r_{ji}^2}-\frac{\bm r_{\parallel}}{\rho_{ji}r_{ij}^2}
\end{align}
where $\bm r_{\parallel}$ is the component of $L_j(\bm n)\bm n$ parallel to the dividing facet $S_{ij}$ between cells $i$ and $j$. One gets the following formula :
\begin{align}
  \bm\nab_i e_j&=-\frac{1}{r_{ij}}\int_{S_{ij}}{\rm d}S (\bm r_{ij}\rho_{ij}+\bm r_{\parallel})(r_{ij}^2\rho_{ji}^2+r_{\parallel}^2+R^2-R_j^2)
\end{align}
Using the global translation invariance, one can write
$  -\bm\nab_i E_p=-\frac{\ga}{2}\sum_{j\neq i} [\bm\nab_i e_j-\bm\nab_j e_i]$.
From $r_{ij}^2\rho_{ji}^2-R_j^2=r_{ij}^2\rho_{ij}^2-R_i^2$, one sees that the components proportional to $\bm r_{\parallel}$ vanish, and we can write
\begin{align}
  -\bm\nab_i E_p&=\frac{\ga}{2}\sum_{j\neq i} \frac{\bm r_{ij}}{r_{ij}}\int_{S_{ij}} {\rm d}S \left(\intvide r_{ij}^2\rho_{ij}\rho_{ji}+r_{\parallel}^2+R^2-\rho_{ij}R_j^2-\rho_{ji}R_i^2\right)\\
  &=\frac{\ga}{2}\oint_{\pa v_i} {\rm d}\bm S\left(\rho_{ij}^2r_{ij}^2+r_{\parallel}^2+R^2-R_i^2\right)
\end{align}
The integral involving $R^2-R_i^2$ yields zero, and one recognizes that the remaining is the surface integral of the function $r^2$, whence $\bm F_i=\frac{\ga}{2}\int_{v_i} {\rm d}^3r \bm\nab (r^2)=\ga\int_{v_i}{\rm d}^3 r \bm r=\ga\bm\tau_i$, i.e. the force is proportional to the cell polarization. In particular, the total polarization $\sum_i\bm \tau_i$ is zero. It is worth noting that this property is intimately related to the Laguerre-Voronoi generalization of the Voronoi tessellation, and that any other generalized tessellation with natural radii would {\em fail} to conserve the total polarization.

\section{Pressure and stress tensor}

\subsection{Equation of state}\label{app:pressure}
For $P_eV=-5U_e/3+(2NT\xi/3)(\pa \phi/\pa\xi)$, we give here  a microscopic formula allowing a numerical computation of the excess pressure. It is convenient to define respectively the ``main'' $E_p^{(m)}$ and ``additional'' $E_p^{(a)}=E_p-E_p^{(m)}$ parts (with some arbitrariness in the denomination) of the potential by
\begin{align}
  E_p^{(m)}&=\frac{\ga}{2}\sum_i\int_{v_i} {\rm d}^3\bm{r}\ r^2\label{Epm}\\
  E_p^{(a)}&=\frac{\ga}{2}\sum_i\int_{v_i} {\rm d}^3\bm{r}\ (R^2-R_i^2)=\frac{\ga}{2}\sum_i v_i(R^2-R_i^2)\label{Epa}
\end{align}
We have using standard arguments:
\begin{align}
  NT\dpar{\phi}{\xi}{T,V,N,N_1}&=\left\lan\dpar{E_p}{\xi}{\bm r}\right\ran
\end{align}
where the last derivative assumes fixed positions of particles in the formula \myref{poten}. A computation along lines similar to that of the preceding Appendix gives
\begin{align}
\dpar{E_p}{\xi}{\bm r}&=E_p^{(a)}/\xi=\frac{\ga v^{2/3}}{2}(V_2\al_1-V_1\al_2)\label{dixneuf}
\end{align}
where $V_a=\sum_{i/R_i=R_a} v_i$ and we recall that $\al_2=1-\al_1=N_2/N$. As a result, from \myref{Pev}, we have for the average potential energy and pressure the following relations:
\begin{align}
  \lan E_p\ran&=\lan E_p^{(m)}\ran+\lan E_p^{(a)}\ran\\
  Pv&=T-\frac{5}{3}\frac{\lan E_p^{(m)}\ran}{N}-\frac{\lan E_p^{(a)}\ran}{N}\label{stateeq}
\end{align}

\subsection{Stress tensor, bulk and shear moduli}\label{app:stress}

\begin{figure}[h]
  \centering
  \includegraphics[width=0.5\textwidth]{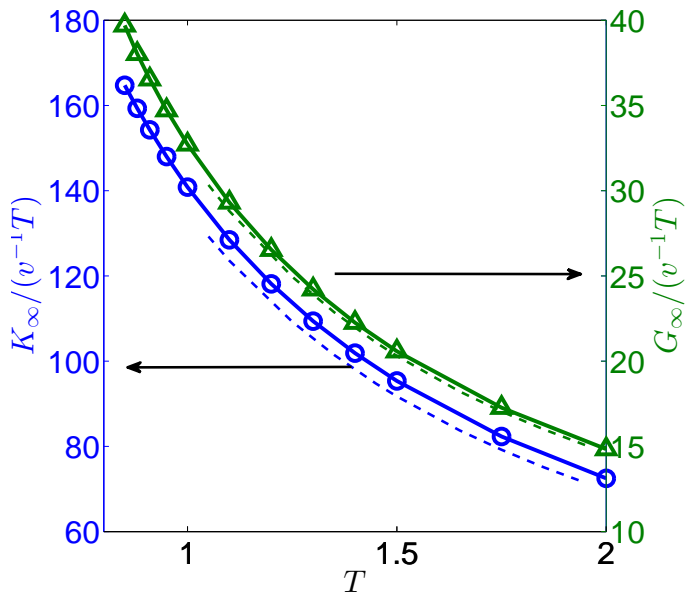}
  \caption{Variations of the bulk ($K_\infty$, blue circles) and shear ($G_\infty$, green triangles) moduli, normalized by the ideal gas shear modulus $G_{\infty}^{\rm id}=T/v$,   with temperature. Dashed line : corresponding quantities for the monodisperse Voronoi fluid.}
  \label{GandK}
\end{figure}

We show here the (nontrivial) fact that the stress tensor is given by the classical relation \myref{classicalstress} with \myref{Fji} as the pair decomposition.
Actually, for an arbitrary potential energy under periodic boundary conditions, the correct expression for the $zz$ component of the stress tensor is \cite{papersound,LouwerseBaerends}
\begin{align}
  \sig_{zz}^{\rm e}=-L_z\frac{\pa E_p}{\pa L_z}-\sum_jz_j\frac{\pa E_p}{\pa z_j}
\end{align}
Using the translation invariance, the right hand side of this expression is readily transformed into $-\sum_{i,j\neq i} \tilde{z}_{ij}(\pa e_i/\pa \tilde{z}_{ij})$ where the tilde stands for the minimum image convention and  $e_i=\frac{\ga}{2}\int_{v_i} {\rm d}^3\bm r(r^2-R_i^2+R^2)$ is the ``local'' energy, which is an implicit function of the $(\tilde{\bm r}_{ik})_{k\neq i}$.
From \myref{ei}, one gets readily (we omit the tilde over distances, but the minimum image convention is implied everywhere)
\begin{align}
  \frac{\pa e_i}{\pa z_{ij}}&=-\frac{\ga}{2r_{ij}}\int_{S_{ij}}dS [r^2+R^2-R_i^2]z'\label{trenta}
\end{align}
where $\bm r$ (resp. $\bm r'$) is a running vector, starting from particle $i$ (resp. $j$) and spanning the dividing surface $S_{ij}$ between cells of $i$ and $j$ (if these cells are not contiguous, the derivative is obviously zero). Notice that $r\neq r'$ if $R_i\neq R_j$. We have subsequently :
\begin{align}
  -\sum_{i,j\neq i} {z}_{ij}\frac{\pa e_i}{\pa {z}_{ij}}&=\frac{\ga}{4}\sum_{i,j\neq i}\frac{z_{ij}}{r_{ij}}\int dS \left\{ [r^2+R^2-R_i^2]z'-[{r'}^2+R^2-R_j^2]z\right\}\\
  &=\frac{\ga}{4}\sum_{i,j\neq i}\frac{z_{ij}}{r_{ij}}\int dS \left\{ [\rho_{ij}^2r_{ij}^2+r_{\parallel}^2+R^2-R_i^2](\rho_{ji}z_{ji}+z_{\parallel})-[\rho_{ji}^2r_{ij}^2+r_\parallel^2+R^2-R_j^2](\rho_{ij}z_{ij}+z_{\parallel})\right\}\\
  &=-\frac{\ga}{4}\sum_{i,j\neq i}\frac{z_{ij}^2}{r_{ij}}\int dS \left\{ r_{ij}^2\rho_{ij}\rho_{ji}+r_\parallel^2+R^2-R_i^2\rho_{ji}-R_j^2\rho_{ij}\right\}\\
  &=-\frac{\ga}{4}\sum_{i,j\neq i}\frac{z_{ij}^2}{r_{ij}}\int dS \left\{ r^2-R_i^2+R^2\right\}=\frac{1}{2}\sum_{i,j\neq i}z_{ji}F_{j\rightarrow i,z}\label{ohoh}
\end{align}
where $\bm r_\parallel=\bm r-\rho_{ij}\bm r_{ij}=\bm r'-\rho_{ji}\bm r_{ji}$ is the component of both $\bm r$ and $\bm r'$ parallel to the facet $S_{ij}$. A quite remarkable and useful explicit expression for $\sig_{zz}^{\rm e}$ can be obtained: If, in \myref{trenta}, $z'$ is replaced by $z_{ji}+z$, one gets, using also \myref{ohoh}, that
\begin{align}
  \sig_{zz}^{\rm e}&=-\frac{\ga}{2}\sum_i \oint_{\pa v_i} {\rm d}\bm S\cdot \bm  e_z \{r^2+R^2-R_i^2\} (\bm r\cdot\bm e_z)\label{trentacinque}
\end{align}

\bigskip

The infinite-frequency bulk modulus is defined by $K_\infty=\chi_T^{-1}+(9VT)^{-1}\lan \de[\sig_{xx}+\sig_{yy}+\sig_{zz}]^2\ran$, where $\chi_T$ is the isothermal compressibility \cite{ZwanzigMountain}. We have from \myref{trentacinque} (using the GO theorem):
\begin{align}
  \sig^{\rm e}_{xx}+\sig^{\rm e}_{yy}+\sig^{\rm e}_{zz}&=-\frac{\ga}{2}\sum_{i} \int_{ v_i} {\rm d}^3r[5r^2+3(R^2-R_i^2)]=-5E_p^{(m)}-3E_p^{(a)}
\end{align}
This result is checked by taking the average of the both sides. We recover the already known relation $3P_ev=-5\lan E_p^{(m)}\ran-3\lan E_p^{(a)}\ran$, where $P_e$ is the excess pressure (see \myref{stateeq}). Taking into account the kinetic part of the stress leads to
\begin{align}
  K_\infty-\chi_T^{-1}&=\frac{2T}{3v}+\frac{1}{VT}\left\lan \de\left(\frac{5}{3}E_p^{( m)}+E_p^{( a)}\right)^2\right\ran
\end{align}
If one considers an infinitesimal triaxial compression of the fluid, one gets a formula for $\chi_T^{-1}$ which compared to the preceding expression, yields the exact relation,
\begin{align}
  K_\infty&=\frac{5T}{3v}+\frac{10}{9V}\lan E_p^{(m)}\ran
\end{align}

\bigskip

In a fluid, the thermal equilibrium is isotropic. As a result, one can show \cite{Balucani} that
$\lan \de\sig_{zz}^2\ran=VT[K_\infty -\chi_T^{-1}+\frac 43G_\infty]$, where the shear modulus $G_\infty=(VT)^{-1}\lan \sig_{xy}^2\ran$ is associated to the nondiagonal element fluctuations of the shear stress. As shown in \cite{papersound}, a {\em non-fluctuating} expression for $G_\infty$ can be written for a large system, which is 
\begin{align}
  G_\infty&=T/v+\frac{1}{2V}\left\lan \sum_{i,j\neq i}\frac{\pa F_{j,x}}{\pa x_i}y_{ij}^2\right\ran\label{46}
\end{align}
On using again the isotropy properties of the liquid, we get a generalized Cauchy relation \cite{ZwanzigMountain}, valid whichever the interactions between particles, provided they are short-ranged \cite{papersound} :
\begin{align}
  \frac{5}{3} G_\infty+\demi K_\infty=\frac{5}{2}\frac{T}{v}+\frac{1}{6V}\left\lan \sum_{i<j} r_{ij}^2\left(\frac{\pa F_{j,x}}{\pa x_i}+\frac{\pa F_{j,y}}{\pa y_i}+\frac{\pa F_{j,z}}{\pa z_i}\right)\right\ran\label{toptropcool}
\end{align}
This relation gives the well-known one \cite{ZwanzigMountain} $\frac 53 G_\infty=K_\infty-2(P-T/v)$ if pair interactions are considered. If not, it is the most isotropic expression linking $G_\infty$ and $K_\infty$ and valid under any circumstances within a fluid.

\medskip

We now use this expression to give an explicit expression, not involving a variance of stress components, for $G_\infty$ for our binary Voronoi fluid.
The Hessian matrix element $\pa^2 E_p/(\pa r_{i,\al}\pa r_{j,\be})=-\pa F_{i,\al}/\pa  r_{j,\be}$ can be simply calculated along lines very similar to those in Appendix \ref{app:forces}. The result is either 0 if $i$ and $j$ are not neighbours, otherwise
\begin{align}
  \frac{\pa F_{i,\be}}{\pa r_{j,\al}}&=-\frac{\ga}{r_{ij}}\int_{S_{ij}} {\rm d}S\ r_{\be}r'_{\al}\label{hesstens}
\end{align}
where, as before, $\bm r$ (resp. $\bm r'$) is a vector starting from $i$ (resp. $j$) and spanning the dividing surface $S_{ij}$; These vectors are related by $\bm r'=\bm r_{ji}+\bm r$. Notice that $\pa F_{i,\be}/\pa r_{i,\al}$ is not required here, but could be obtained from the translation invariance : $\pa F_{i,\be}/\pa r_{i,\al}=-\sum_{j\neq i}\pa F_{i,\be}/\pa r_{j,\al}$.
From \myref{toptropcool}, we get ($S_{ij}$ is the area of the facet common to the cells $i$ and $j$, if it exists, or zero otherwise)
\begin{align}
  G_\infty&=T/v-\frac 4{3V}\lan E_p^{(m)}\ran+\frac{\ga}{20V}\left\lan\sum_{i<j}\left[1+\frac{(R_i^2-R_j^2)^2}{r_{ij}^4}\right]S_{ij}r_{ij}^3\right\ran
\end{align}
This result shows that $G_\infty$, when expressed as an expression not involving a variance of stress, comes as a {\em difference} between two positive terms. It is interesting to note that within this expression, the positivity of $G_\infty-T/v$ is far from obvious. One can also give another form to $G_\infty$ by splitting the running vectors $\bm r$ going from $i$ to the facets into $\bm r=\rho_{ij}\bm r_{ij}+\bm r_\parallel$, where $\bm r_{\parallel}$ is the component parallel to the facet :
\begin{align}
  G_\infty&=\frac Tv+\frac{\ga}{60}\left\lan\sum_{i<j} r_{ij}\left(S_{ij}r_{ij}^2-3S_{ij}r_{ij}^{-2}(R_i^2-R_j^2)^2-8\int_{S_{ij}}{\rm d}S\ r_\parallel^2\right)\right\ran
\end{align}
We see that $G_\infty-T/v$ is the sum of one positive term involving globally the distance $r_{ij}$ and the surface of the facet, corrected by two negative terms, one  involving explicitely the polydispersity ($\propto {\xi}^2$ which is quite small if the polydispersity  factor $\xi$ is small), the other accounting for geometrical details of the facets. Very roughly, the positivity of $G_\infty$ (for small polydispersity) can be understood by the fact that one has typically $\int_{S_{ij}} dS r_\parallel^2\sim \frac 12 S_{ij}r_0^2$,  where $r_0$ is the typical radius of a facet. Therefore the positivity of $G_\infty-T/v$ is roughly checked if $r_{ij}>2r_0$. A typical cell has a radius $\sim r_{ij}/2$ and $\sim 12-14$ faces, therefore the typical aperture of the solid angle under which a face is seen is $\te_0$ such that $1-\cos(\te_0)\sim 1/7$, thus $\te_0\sim \pi/6$. As a result $r_0\sim \sin(\te_0)r_{ij}/2\sim r_{ij}/4$, thus the condition $r_{ij}>2r_0$ is amply fulfilled for nearly spherical Voronoi cells. But this quick argument must not hide the fact that the positivity of $G_\infty-T/v$ is always guaranteed by \myref{46}, whatever the polydispersity and the temperature.

\section{Evolution of the rdf maxima with polydispersity}\label{app:rdfmax}

Let us term $r_{ab}^*$ ($a,b\in\{1,2\}$) the location of the first maximum of $g_{ab}(r)$ defined in section \ref{sectiongder}. When $\xi\rightarrow0$, the monodisperse fluid is recovered and $r_{ab}^*=r_0^*$ is independent of $a$ and $b$. For small but nonzero $\xi$, one writes thus $r_{ab}^*=r_0^*+\de r_{ab}^*$, and the purpose of this section is to compute and to compare the $\de r_{ab}^*$ in this limit. The discussion is not limited to the particular case $\al_1=1/2$ considered otherwise in this work.

\medskip

\begin{figure}[h]
  \centering
  \includegraphics[width=0.5\textwidth]{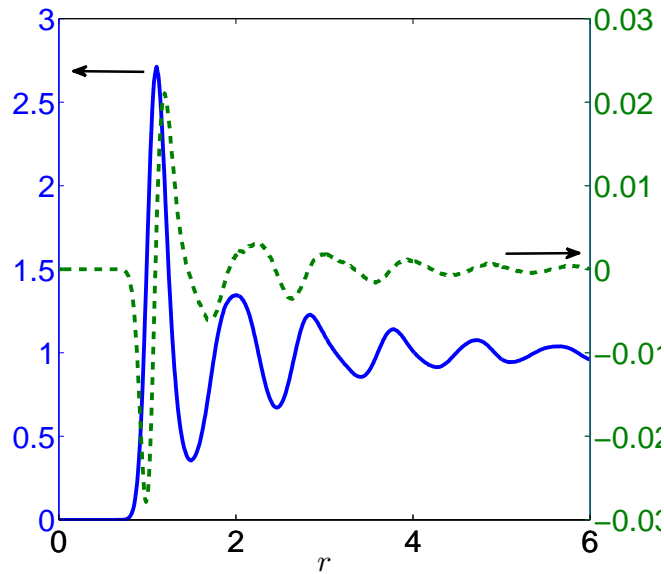}
  \caption{Solid blue : $g(r)$. Dashed green : $\sum_{j/j\neq i}\lan \de(\bm r-\bm r_{ij})[v_i-v]\ran$. Monodisperse fluid at $T=1.05$. This figure shows that at each maximum (resp. minimum) of $g(r)$ the dashed curve has a positive (resp. negative) slope.}
  \label{gv}
\end{figure}

The  definition of $r_{ab}^*$ allows to write, at the lowest order :
\begin{align}
  g_{ab}'(r_{ab}^*)=0 \Rightarrow \de r_{ab}^*=-\frac{\de g_{ab}'(r_0^*)}{g''_0(r_0^*)}
\end{align}
where $g_0(r)$ is the rdf of the monodisperse system and $\de g_{ab}=g_{ab}-g_0$, i.e. the variation of $g_{ab}$ when the polydispersity parameter is switched from 0 to $\xi$. From the microscopic definition of $g_{ab}(r)$ and the formula \myref{Eprime} $\pa E_p/\pa \xi=\frac{\ga v^{2/3}}{2}(V\al_1-V_1)$, one gets
\begin{align}
  \de g_{ab}(r)&=\frac{\be \ga v^{2/3} V\xi}{2}\lan \de(\bm r-\bm r_{ij})[V_1-\lan V_1\ran_0]\ran_0
\end{align}
with the important precision that $i$ (resp. $j$) is a particle of type $a$ (resp. $b$). The subscript ``$0$'' stands for a canonical average with the monodisperse energy $\xi=0$. This formula is puzzling at first sight, since a reference to type ``1'' or type ``2'' is made within an average over monodisperse configurations where the particles have so to speak lost their type. This remark allows to rewrite differently (and more appealingly) this formula, this rewriting depends on whether $(a,b)=(1,1),(2,2)$ or $(1,2)=(2,1)$. Let us consider the first case. We have
\begin{align}
  \lan \de(\bm r-\bm r_{ij})[V_1-\lan V_1\ran_0]\ran_0&=2  \lan \de(\bm r-\bm r_{ij})[v_i-v]\ran_0+(N_1-2)\lan \de(\bm r-\bm r_{ij})[v_k-v]\ran_0
\end{align}
where $k\not\in\{i,j\}$. On using $\lan \de(\bm r-\bm r_{ij})[V-\lan V\ran_0]\ran_0=0$, the last average (with $k$) can be transformed into $-2\lan \de(\bm r-\bm r_{ij})[v_i-v]\ran_0/(N-2)$, and altogether we have
\begin{align}
  \de r_{11}^*&=\frac{\xi\be v^{5/3} \ga Q}{|g''_0(r^*)|}\al_2\\
  Q&=\frac{d}{dr}\left[\sum_{j/j\neq i}\lan \de(\bm r-\bm r_{ij})[v_i-v]\ran_0\right]_{(r=r^*)}
\end{align}
(we used the fact that $g''_0(r^*)<0$). On the other hand,  similar computations yield
\begin{align}
  \de r_{22}^*&=\frac{\xi\be v^{5/3} \ga Q}{|g''_0(r^*)|}(-\al_1)\\
  \de r_{12}^*&=\frac{\xi\be v^{5/3} \ga Q}{|g''_0(r^*)|}\frac{\al_2-\al_1}{2}
\end{align}
These results  show the following surprising results : In the limit of small values of $\xi$, one has that
\begin{itemize}
\item $\de r^*_{11}-\de r^*_{12}=\de r^*_{12}-\de r^*_{22}\equiv \De$
\item $\De=(\xi\be \ga  v^{5/3}Q/(2|g''_0(r^*)|)$ is independent of $\al_1$
  \item These results apply to {\em any} extremum of the rdf.
\end{itemize}
In other words, the first maximum of $g_{12}(r)$ is always in the middle of those of $g_{11}$ and $g_{22}$. Besides, the width of the spread of these three maxima is $2\De$, {\em independent} of the composition of the fluid ! Moreover, this result is valid for all triplets coming from any extremum (minimum or maximum) of $g(r)$, the spread being however dependent on the extremum considered.

A remark is in order concerning the sign of $Q$: On physical grounds, one expects $Q>0$ if a maximum of $g(r)$ is considered, in order to have $\de r_{22}^*<0<\de r_{11}^*$. We observe also in fig. \ref{gder} that conversely, one must have $Q<0$ if a minimum of $g(r)$ is considered (since one has $g''>0$ here). That it is indeed the case can be seen in fig. \ref{gv}, where $g(r)$ and $\sum_{j/j\neq i}\lan \de(\bm r-\bm r_{ij})[v_i-v]\ran$
for the monodisperse fluid at $T=1.05$ are superimposed. It would be interesting to go beyond this mere observation, but to {\em prove} that it is always the case  seems a difficult task.

Finally, it is worth noticing that a similar expansion performed around the extrema of the structure factor yields exactly the same conclusions for the relative placing of the extrema of $S_{11}(k)$, $S_{22}(k)$ and $S_{12}(k)$ (the maximum of $S_{11}(k)$ being of course shifted leftward).

\bibliography{toutvenant,voronoi,mespapiers}

\begin{thebibliography}{25}%
\makeatletter
\providecommand \@ifxundefined [1]{%
 \@ifx{#1\undefined}
}%
\providecommand \@ifnum [1]{%
 \ifnum #1\expandafter \@firstoftwo
 \else \expandafter \@secondoftwo
 \fi
}%
\providecommand \@ifx [1]{%
 \ifx #1\expandafter \@firstoftwo
 \else \expandafter \@secondoftwo
 \fi
}%
\providecommand \natexlab [1]{#1}%
\providecommand \enquote  [1]{``#1''}%
\providecommand \bibnamefont  [1]{#1}%
\providecommand \bibfnamefont [1]{#1}%
\providecommand \citenamefont [1]{#1}%
\providecommand \href@noop [0]{\@secondoftwo}%
\providecommand \href [0]{\begingroup \@sanitize@url \@href}%
\providecommand \@href[1]{\@@startlink{#1}\@@href}%
\providecommand \@@href[1]{\endgroup#1\@@endlink}%
\providecommand \@sanitize@url [0]{\catcode `\\12\catcode `\$12\catcode
  `\&12\catcode `\#12\catcode `\^12\catcode `\_12\catcode `\%12\relax}%
\providecommand \@@startlink[1]{}%
\providecommand \@@endlink[0]{}%
\providecommand \url  [0]{\begingroup\@sanitize@url \@url }%
\providecommand \@url [1]{\endgroup\@href {#1}{\urlprefix }}%
\providecommand \urlprefix  [0]{URL }%
\providecommand \Eprint [0]{\href }%
\providecommand \doibase [0]{http://dx.doi.org/}%
\providecommand \selectlanguage [0]{\@gobble}%
\providecommand \bibinfo  [0]{\@secondoftwo}%
\providecommand \bibfield  [0]{\@secondoftwo}%
\providecommand \translation [1]{[#1]}%
\providecommand \BibitemOpen [0]{}%
\providecommand \bibitemStop [0]{}%
\providecommand \bibitemNoStop [0]{.\EOS\space}%
\providecommand \EOS [0]{\spacefactor3000\relax}%
\providecommand \BibitemShut  [1]{\csname bibitem#1\endcsname}%
\let\auto@bib@innerbib\@empty
\bibitem [{\citenamefont {Cavagna}(2009)}]{CavagnaReport}%
  \BibitemOpen
  \bibfield  {author} {\bibinfo {author} {\bibfnamefont {A.}~\bibnamefont
  {Cavagna}},\ }\href@noop {} {\bibfield  {journal} {\bibinfo  {journal}
  {Physics Reports}\ }\textbf {\bibinfo {volume} {476}},\ \bibinfo {pages} {51}
  (\bibinfo {year} {2009})}\BibitemShut {NoStop}%
\bibitem [{\citenamefont {Leuzzi}\ and\ \citenamefont
  {Th.~M}(2007)}]{NieuwenhuizenBook}%
  \BibitemOpen
  \bibfield  {author} {\bibinfo {author} {\bibfnamefont {L.}~\bibnamefont
  {Leuzzi}}\ and\ \bibinfo {author} {\bibfnamefont {N.}~\bibnamefont {Th.~M}},\
  }\href@noop {} {\emph {\bibinfo {title} {Thermodynamics of the Glassy
  State}}}\ (\bibinfo  {publisher} {CRC Press},\ \bibinfo {year}
  {2007})\BibitemShut {NoStop}%
\bibitem [{\citenamefont {Berthier}\ and\ \citenamefont
  {Biroli}(2011)}]{BerthierBiroliRMP}%
  \BibitemOpen
  \bibfield  {author} {\bibinfo {author} {\bibfnamefont {L.}~\bibnamefont
  {Berthier}}\ and\ \bibinfo {author} {\bibfnamefont {G.}~\bibnamefont
  {Biroli}},\ }\href {\doibase 10.1103/RevModPhys.83.587} {\bibfield  {journal}
  {\bibinfo  {journal} {Rev. Mod. Phys.}\ }\textbf {\bibinfo {volume} {83}},\
  \bibinfo {pages} {587} (\bibinfo {year} {2011})}\BibitemShut {NoStop}%
\bibitem [{\citenamefont {{Ruscher, C.}}\ \emph {et~al.}(2015)\citenamefont
  {{Ruscher, C.}}, \citenamefont {{Baschnagel, J.}},\ and\ \citenamefont
  {{Farago, J.}}}]{voronoiliquid}%
  \BibitemOpen
  \bibfield  {author} {\bibinfo {author} {\bibnamefont {{Ruscher, C.}}},
  \bibinfo {author} {\bibnamefont {{Baschnagel, J.}}}, \ and\ \bibinfo {author}
  {\bibnamefont {{Farago, J.}}},\ }\href {\doibase 10.1209/0295-5075/112/66003}
  {\bibfield  {journal} {\bibinfo  {journal} {EPL}\ }\textbf {\bibinfo {volume}
  {112}},\ \bibinfo {pages} {66003} (\bibinfo {year} {2015})}\BibitemShut
  {NoStop}%
\bibitem [{\citenamefont {Ruscher}\ \emph {et~al.}(2017)\citenamefont
  {Ruscher}, \citenamefont {Semenov}, \citenamefont {Baschnagel},\ and\
  \citenamefont {Farago}}]{papersound}%
  \BibitemOpen
  \bibfield  {author} {\bibinfo {author} {\bibfnamefont {C.}~\bibnamefont
  {Ruscher}}, \bibinfo {author} {\bibfnamefont {A.~N.}\ \bibnamefont
  {Semenov}}, \bibinfo {author} {\bibfnamefont {J.}~\bibnamefont {Baschnagel}},
  \ and\ \bibinfo {author} {\bibfnamefont {J.}~\bibnamefont {Farago}},\
  }\href@noop {} {\bibfield  {journal} {\bibinfo  {journal} {The Journal of
  Chemical Physics}\ }\textbf {\bibinfo {volume} {146}},\ \bibinfo {pages}
  {144502} (\bibinfo {year} {2017})}\BibitemShut {NoStop}%
\bibitem [{\citenamefont {Balucani}\ and\ \citenamefont
  {Zoppi}(1995)}]{Balucani}%
  \BibitemOpen
  \bibfield  {author} {\bibinfo {author} {\bibfnamefont {U.}~\bibnamefont
  {Balucani}}\ and\ \bibinfo {author} {\bibfnamefont {M.}~\bibnamefont
  {Zoppi}},\ }\href@noop {} {\emph {\bibinfo {title} {Dynamics of the Liquid
  State}}}\ (\bibinfo  {publisher} {Oxford University Press},\ \bibinfo {year}
  {1995})\BibitemShut {NoStop}%
\bibitem [{\citenamefont {Plimpton}(1995)}]{Plimpton}%
  \BibitemOpen
  \bibfield  {author} {\bibinfo {author} {\bibfnamefont {S.}~\bibnamefont
  {Plimpton}},\ }\href {\doibase http://dx.doi.org/10.1006/jcph.1995.1039}
  {\bibfield  {journal} {\bibinfo  {journal} {Journal of Computational
  Physics}\ }\textbf {\bibinfo {volume} {117}},\ \bibinfo {pages} {1 }
  (\bibinfo {year} {1995})}\BibitemShut {NoStop}%
\bibitem [{\citenamefont {Rycroft}(2009)}]{voro++}%
  \BibitemOpen
  \bibfield  {author} {\bibinfo {author} {\bibfnamefont {C.~H.}\ \bibnamefont
  {Rycroft}},\ }\href@noop {} {\bibfield  {journal} {\bibinfo  {journal}
  {Chaos}\ }\textbf {\bibinfo {volume} {19}},\ \bibinfo {pages} {041111}
  (\bibinfo {year} {2009})}\BibitemShut {NoStop}%
\bibitem [{\citenamefont {Okabe}\ \emph {et~al.}(2000)\citenamefont {Okabe},
  \citenamefont {Boots}, \citenamefont {Sugihara},\ and\ \citenamefont
  {Nok~Chiu}}]{SpatialTessellations}%
  \BibitemOpen
  \bibfield  {author} {\bibinfo {author} {\bibfnamefont {A.}~\bibnamefont
  {Okabe}}, \bibinfo {author} {\bibfnamefont {B.}~\bibnamefont {Boots}},
  \bibinfo {author} {\bibfnamefont {K.}~\bibnamefont {Sugihara}}, \ and\
  \bibinfo {author} {\bibfnamefont {S.}~\bibnamefont {Nok~Chiu}},\ }\href@noop
  {} {\emph {\bibinfo {title} {Spatial Tessellations: Concepts and Applications
  of Voronoi Diagrams}}}\ (\bibinfo  {publisher} {Wiley},\ \bibinfo {year}
  {2000})\BibitemShut {NoStop}%
\bibitem [{\citenamefont {Callen}(1985)}]{callen}%
  \BibitemOpen
  \bibfield  {author} {\bibinfo {author} {\bibfnamefont {H.}~\bibnamefont
  {Callen}},\ }\href@noop {} {\emph {\bibinfo {title} {Thermodynamics and an
  Introduction to Thermostatistics}}}\ (\bibinfo  {publisher} {Wiley, New
  York},\ \bibinfo {year} {1985})\BibitemShut {NoStop}%
\bibitem [{\citenamefont {Starr}\ \emph {et~al.}(2002)\citenamefont {Starr},
  \citenamefont {Sastry}, \citenamefont {Douglas},\ and\ \citenamefont
  {Glotzer}}]{StarrSastryDouglasGlotzer}%
  \BibitemOpen
  \bibfield  {author} {\bibinfo {author} {\bibfnamefont {F.~W.}\ \bibnamefont
  {Starr}}, \bibinfo {author} {\bibfnamefont {S.}~\bibnamefont {Sastry}},
  \bibinfo {author} {\bibfnamefont {J.~F.}\ \bibnamefont {Douglas}}, \ and\
  \bibinfo {author} {\bibfnamefont {S.~C.}\ \bibnamefont {Glotzer}},\ }\href
  {\doibase 10.1103/PhysRevLett.89.125501} {\bibfield  {journal} {\bibinfo
  {journal} {Phys. Rev. Lett.}\ }\textbf {\bibinfo {volume} {89}},\ \bibinfo
  {pages} {125501} (\bibinfo {year} {2002})}\BibitemShut {NoStop}%
\bibitem [{Note1()}]{Note1}%
  \BibitemOpen
  \bibinfo {note} {Actually, one could even go one step further, since it can
  be shown that the facets $S_{ij}$ are second order polynomials of $\xi $. The
  expressions become howevcer involved.}\BibitemShut {Stop}%
\bibitem [{\citenamefont {Heuer}(2008)}]{HeuerReview}%
  \BibitemOpen
  \bibfield  {author} {\bibinfo {author} {\bibfnamefont {A.}~\bibnamefont
  {Heuer}},\ }\href {http://stacks.iop.org/0953-8984/20/i=37/a=373101}
  {\bibfield  {journal} {\bibinfo  {journal} {Journal of Physics: Condensed
  Matter}\ }\textbf {\bibinfo {volume} {20}},\ \bibinfo {pages} {373101}
  (\bibinfo {year} {2008})}\BibitemShut {NoStop}%
\bibitem [{\citenamefont {Hansen}\ and\ \citenamefont
  {McDonalds}(2006)}]{Hansen}%
  \BibitemOpen
  \bibfield  {author} {\bibinfo {author} {\bibfnamefont {J.~P.}\ \bibnamefont
  {Hansen}}\ and\ \bibinfo {author} {\bibfnamefont {I.~R.}\ \bibnamefont
  {McDonalds}},\ }\href@noop {} {\emph {\bibinfo {title} {Theory of Simple
  Liquids}}},\ \bibinfo {edition} {3rd}\ ed.\ (\bibinfo  {publisher} {Academic
  Press},\ \bibinfo {year} {2006})\BibitemShut {NoStop}%
\bibitem [{\citenamefont {Verlet}(1968)}]{Verlet165}%
  \BibitemOpen
  \bibfield  {author} {\bibinfo {author} {\bibfnamefont {L.}~\bibnamefont
  {Verlet}},\ }\href@noop {} {\bibfield  {journal} {\bibinfo  {journal} {Phys.
  Rev.}\ }\textbf {\bibinfo {volume} {165}},\ \bibinfo {pages} {201} (\bibinfo
  {year} {1968})}\BibitemShut {NoStop}%
\bibitem [{\citenamefont {Admal}\ and\ \citenamefont
  {Tadmor}(2011)}]{admaltadmor}%
  \BibitemOpen
  \bibfield  {author} {\bibinfo {author} {\bibfnamefont {N.~C.}\ \bibnamefont
  {Admal}}\ and\ \bibinfo {author} {\bibfnamefont {E.~B.}\ \bibnamefont
  {Tadmor}},\ }\href {\doibase http://dx.doi.org/10.1063/1.3582905} {\bibfield
  {journal} {\bibinfo  {journal} {The Journal of Chemical Physics}\ }\textbf
  {\bibinfo {volume} {134}},\ \bibinfo {eid} {184106} (\bibinfo {year}
  {2011})}\BibitemShut {NoStop}%
\bibitem [{\citenamefont {Broderix}\ \emph {et~al.}(2000)\citenamefont
  {Broderix}, \citenamefont {Bhattacharya}, \citenamefont {Cavagna},
  \citenamefont {Zippelius},\ and\ \citenamefont {Giardina}}]{BroderixSaddles}%
  \BibitemOpen
  \bibfield  {author} {\bibinfo {author} {\bibfnamefont {K.}~\bibnamefont
  {Broderix}}, \bibinfo {author} {\bibfnamefont {K.~K.}\ \bibnamefont
  {Bhattacharya}}, \bibinfo {author} {\bibfnamefont {A.}~\bibnamefont
  {Cavagna}}, \bibinfo {author} {\bibfnamefont {A.}~\bibnamefont {Zippelius}},
  \ and\ \bibinfo {author} {\bibfnamefont {I.}~\bibnamefont {Giardina}},\
  }\href {\doibase 10.1103/PhysRevLett.85.5360} {\bibfield  {journal} {\bibinfo
   {journal} {Phys. Rev. Lett.}\ }\textbf {\bibinfo {volume} {85}},\ \bibinfo
  {pages} {5360} (\bibinfo {year} {2000})}\BibitemShut {NoStop}%
\bibitem [{\citenamefont {Angelani}\ \emph {et~al.}(2000)\citenamefont
  {Angelani}, \citenamefont {Di~Leonardo}, \citenamefont {Ruocco},
  \citenamefont {Scala},\ and\ \citenamefont {Sciortino}}]{AngelaniSaddles}%
  \BibitemOpen
  \bibfield  {author} {\bibinfo {author} {\bibfnamefont {L.}~\bibnamefont
  {Angelani}}, \bibinfo {author} {\bibfnamefont {R.}~\bibnamefont
  {Di~Leonardo}}, \bibinfo {author} {\bibfnamefont {G.}~\bibnamefont {Ruocco}},
  \bibinfo {author} {\bibfnamefont {A.}~\bibnamefont {Scala}}, \ and\ \bibinfo
  {author} {\bibfnamefont {F.}~\bibnamefont {Sciortino}},\ }\href {\doibase
  10.1103/PhysRevLett.85.5356} {\bibfield  {journal} {\bibinfo  {journal}
  {Phys. Rev. Lett.}\ }\textbf {\bibinfo {volume} {85}},\ \bibinfo {pages}
  {5356} (\bibinfo {year} {2000})}\BibitemShut {NoStop}%
\bibitem [{\citenamefont {Wahnstr\"om}(1991)}]{Wahnstrom}%
  \BibitemOpen
  \bibfield  {author} {\bibinfo {author} {\bibfnamefont {G.}~\bibnamefont
  {Wahnstr\"om}},\ }\href {\doibase 10.1103/PhysRevA.44.3752} {\bibfield
  {journal} {\bibinfo  {journal} {Phys. Rev. A}\ }\textbf {\bibinfo {volume}
  {44}},\ \bibinfo {pages} {3752} (\bibinfo {year} {1991})}\BibitemShut
  {NoStop}%
\bibitem [{\citenamefont {Sciortino}(2005)}]{sciortinoJSTATMECH}%
  \BibitemOpen
  \bibfield  {author} {\bibinfo {author} {\bibfnamefont {F.}~\bibnamefont
  {Sciortino}},\ }\href@noop {} {\bibfield  {journal} {\bibinfo  {journal}
  {Journal of Statistical Mechanics: Theory and Experiment}\ }\textbf {\bibinfo
  {volume} {2005}},\ \bibinfo {pages} {P05015} (\bibinfo {year}
  {2005})}\BibitemShut {NoStop}%
\bibitem [{\citenamefont {Zaccarelli}\ \emph {et~al.}(2002)\citenamefont
  {Zaccarelli}, \citenamefont {Foffi}, \citenamefont {Gregorio}, \citenamefont
  {Sciortino}, \citenamefont {Tartaglia},\ and\ \citenamefont
  {Dawson}}]{ZFGST}%
  \BibitemOpen
  \bibfield  {author} {\bibinfo {author} {\bibfnamefont {E.}~\bibnamefont
  {Zaccarelli}}, \bibinfo {author} {\bibfnamefont {G.}~\bibnamefont {Foffi}},
  \bibinfo {author} {\bibfnamefont {P.~D.}\ \bibnamefont {Gregorio}}, \bibinfo
  {author} {\bibfnamefont {F.}~\bibnamefont {Sciortino}}, \bibinfo {author}
  {\bibfnamefont {P.}~\bibnamefont {Tartaglia}}, \ and\ \bibinfo {author}
  {\bibfnamefont {K.~A.}\ \bibnamefont {Dawson}},\ }\href@noop {} {\bibfield
  {journal} {\bibinfo  {journal} {Journal of Physics: Condensed Matter}\
  }\textbf {\bibinfo {volume} {14}},\ \bibinfo {pages} {2413} (\bibinfo {year}
  {2002})}\BibitemShut {NoStop}%
\bibitem [{\citenamefont {{Jean Farago}}\ \emph {et~al.}(2014)\citenamefont
  {{Jean Farago}}, \citenamefont {{Alexander Semenov}}, \citenamefont {{Stefan
  Frey}},\ and\ \citenamefont {{Joerg Baschnagel}}}]{epjeVoronoi}%
  \BibitemOpen
  \bibfield  {author} {\bibinfo {author} {\bibnamefont {{Jean Farago}}},
  \bibinfo {author} {\bibnamefont {{Alexander Semenov}}}, \bibinfo {author}
  {\bibnamefont {{Stefan Frey}}}, \ and\ \bibinfo {author} {\bibnamefont
  {{Joerg Baschnagel}}},\ }\href {\doibase 10.1140/epje/i2014-14046-9}
  {\bibfield  {journal} {\bibinfo  {journal} {Eur. Phys. J. E}\ }\textbf
  {\bibinfo {volume} {37}},\ \bibinfo {pages} {46} (\bibinfo {year}
  {2014})}\BibitemShut {NoStop}%
\bibitem [{\citenamefont {Song}\ \emph {et~al.}(2010)\citenamefont {Song},
  \citenamefont {Wang}, \citenamefont {Jin},\ and\ \citenamefont
  {Makse}}]{makse}%
  \BibitemOpen
  \bibfield  {author} {\bibinfo {author} {\bibfnamefont {C.}~\bibnamefont
  {Song}}, \bibinfo {author} {\bibfnamefont {P.}~\bibnamefont {Wang}}, \bibinfo
  {author} {\bibfnamefont {Y.}~\bibnamefont {Jin}}, \ and\ \bibinfo {author}
  {\bibfnamefont {H.~A.}\ \bibnamefont {Makse}},\ }\href {\doibase
  10.1016/j.physa.2010.06.043} {\bibfield  {journal} {\bibinfo  {journal}
  {Physica A: Statistical Mechanics and its Applications}\ }\textbf {\bibinfo
  {volume} {389}},\ \bibinfo {pages} {4497 } (\bibinfo {year}
  {2010})}\BibitemShut {NoStop}%
\bibitem [{\citenamefont {Louwerse}\ and\ \citenamefont
  {Baerends}(2006)}]{LouwerseBaerends}%
  \BibitemOpen
  \bibfield  {author} {\bibinfo {author} {\bibfnamefont {M.~J.}\ \bibnamefont
  {Louwerse}}\ and\ \bibinfo {author} {\bibfnamefont {E.~J.}\ \bibnamefont
  {Baerends}},\ }\href {\doibase
  http://dx.doi.org/10.1016/j.cplett.2006.01.087} {\bibfield  {journal}
  {\bibinfo  {journal} {Chemical Physics Letters}\ }\textbf {\bibinfo {volume}
  {421}},\ \bibinfo {pages} {138 } (\bibinfo {year} {2006})}\BibitemShut
  {NoStop}%
\bibitem [{\citenamefont {Zwanzig}\ and\ \citenamefont
  {Mountain}(1965)}]{ZwanzigMountain}%
  \BibitemOpen
  \bibfield  {author} {\bibinfo {author} {\bibfnamefont {R.}~\bibnamefont
  {Zwanzig}}\ and\ \bibinfo {author} {\bibfnamefont {R.~D.}\ \bibnamefont
  {Mountain}},\ }\href {\doibase http://dx.doi.org/10.1063/1.1696718}
  {\bibfield  {journal} {\bibinfo  {journal} {The Journal of Chemical Physics}\
  }\textbf {\bibinfo {volume} {43}},\ \bibinfo {pages} {4464} (\bibinfo {year}
  {1965})}\BibitemShut {NoStop}%
\end{thebibliography}%

\end{document}